 \documentclass[preprint,review,12pt]{elsarticle}

\usepackage{graphics}
\usepackage{amssymb}
\usepackage{subeqn}

\biboptions{sort&compress}

\journal{}

\begin{document}

\begin{frontmatter}

\title{A lattice Boltzmann model for two-phase flow in porous media}

%% use optional labels to link authors explicitly to addresses:
%% \author[label1,label2]{<author name>}
%% \address[label1]{<address>}
%% \address[label2]{<address>}

\author{Zhenhua Chai\fnref{a,b,c}}\ead{hustczh@hust.edu.cn}
\author{Hong Liang\fnref{d}}
\author{Rui Du\fnref{e}}
\author[a,b,c]{Baochang Shi\corref{cor1}}\cortext[cor1]{Corresponding author at: School of Mathematics and Statistics, Huazhong University of Science and Technology, Wuhan, 430074, China. Tel./fax: +86 27 8754 3231. \hspace*{20pt}}\ead{shibc@hust.edu.cn}
\address[a]{School of Mathematics and Statistics, Huazhong University of Science and Technology, Wuhan, 430074, China}
\address[b]{Hubei Key Laboratory of Engineering Modeling and Scientific Computing, Huazhong University of Science and Technology, Wuhan 430074, China}
\address[c]{State Key Laboratory of Coal Combustion, Huazhong University of Science and Technology, Wuhan 430074, China}
\address[d]{Department of Physics, Hangzhou Dianzi University, Hangzhou 310018, China}
\address[e]{School of Mathematics, Southeast University, Nanjing 210096, China}

\begin{abstract}

In this paper, a lattice Boltzmann (LB) model with double distribution functions is proposed for two-phase flow in porous media where one distribution function is used for pressure governed by the Poisson equation, and the other is applied for saturation evolution described by the convection-diffusion equation with a source term. We first performed a Chapman-Enskog analysis, and show that the macroscopic nonlinear equations for pressure and saturation can be recovered correctly from present LB model. Then in the framework of LB method, we develop a local scheme for pressure gradient or equivalently velocity, which may be more efficient than the nonlocal second-order finite-difference schemes. We also perform some numerical simulations, and the results show that the developed LB model and local scheme for velocity are accurate and also have a second-order convergence rate in space. Finally, compared to the available pore-scale LB models for two-phase flow in porous media, the present LB model has more potential in the study of the large-scale problems.

\end{abstract}

\begin{keyword}
%% keywords here, in the form: keyword \sep keyword
lattice Boltzmann model \sep two-phase flow \sep porous media \sep Chapman-Enskog analysis
%% MSC codes here, in the form: \MSC code \sep code
%% or \MSC[2008] code \sep code (2000 is the default)

\end{keyword}

\end{frontmatter}

%
% Start line numbering here if you want
%
%\linenumbers

% main text
\section{Introduction}

The processes of two-phase flows through porous media are universal and also important in both science and engineering, such as ground water, soil sciences, petroleum engineering, CO$_{2}$ geologic sequestration and fuel cells \cite{Bear1972,Chen2006,Wu2016,Zhao2009}. Because of their wide applications in practice, tremendous research and great efforts have been devoted to the study of two-phase flows in porous media from pore and macroscopic scale levels. At pore-scale level, the two-phase fluid flows can be depicted by the Navier-Stokes and interface capturing equations \cite{Anderson1998,Tryggvason2011}, and through solving these governing equations, one can obtain the detailed information of fluid flows in porous media \cite{Blunt2013,Liu2016}. For this reason, the pore-scale approaches based on Navier-Stokes and interface capturing equations are usually adopted to reveal the physical mechanism and predict some physical properties of two-phase flows through porous media. Due to the complexity of pore structure of porous media and expensive computational cost, however, these pore-scale methods are restricted to a small physical region, and cannot be extended readily to investigate large-scale problems. On the contrary, the macroscopic continuum models for two-phase flows in porous media, developed through appropriate volumetric averaging of the governing equations at pore scale and based on the representative elementary volume \cite{Bear1972}, are most commonly used in the study of large-scale problems \cite{Wu2016}. However, due to the nonlinearity and coupling of these macroscopic continuum models, it is difficult or even impossible to obtain their analytical solutions \cite{Bear1972,Chen2006,Chavent1986}. Fortunately, with the development of computational technology, some numerical approaches, including finite-difference method \cite{Jr1983}, finite-volume method \cite{Enchery2006,Durlofsky2007}, finite-element method \cite{Chen2006,Chavent1986,Lewis1984,Kukreti1989}, operator-splitting method \cite{Cao2011,Chueh2013}, implicit pressure-explicit saturation method \cite{Chen2006,Chen2004} and active-set reduced-space method \cite{Yang2016}, have been proposed to solve the macroscopic continuum models, and also gained a great success in the study of two-phase flows in porous media \cite{Chen2006,Gerritsen2005}. In this work, we will present an alternative, i.e., lattice Boltzmann (LB) method, for two-phase flows in porous media. Compared to above mentioned traditional methods, the LB method has some distinct advantages in implementation of complex boundary conditions and efficiency of parallel computing \cite{Chen1998,Succi2001,Guo2013,Kruger2017}.

The LB method, as a mesoscopic numerical approach, has attained increasing attention in the study of fluid flows and some special physical systems governed by convection-diffusion equations (CDEs) \cite{Chen1998,Succi2001,Guo2013,Kruger2017,Chai2016}. Based on its kinetic nature, the LB method has been extended to investigate the multiphase flows in porous media \cite{Liu2016,Chen2014,Li2016,Xu2017} (see references therein), and has also been viewed as one of the most popular pore-scale approaches for such complicated problems \cite{Blunt2013,Bultreys2016}. Although these available works based on LB method can be used to explore the physical mechanism and predict some physical properties from pore-scale level, they cannot be readily applied for large-scale problems since the computational cost is too expensive, as stated previously. The aim of the present work is to fill the gap through developing a LB model for two-phase flows in porous media from macroscopic scale level.

The rest of the paper is organized as follows. In Section 2, the mathematical model for two-phase flows in porous media is first introduced, then the LB model with double distribution functions is proposed in Section 3. In Section 4, the present LB model is tested through some benchmark problems, and finally, some conclusions are given in Section 5.

\section{Mathematical model for two-phase flow in porous media}

At macroscopic scale level, the mass conservation equation of incompressible two-phase fluid flows in porous media can be written as \cite{Bear1972,Chen2006}
\begin{equation}\label{eq2-1}
\phi\frac{\partial (\rho_{\alpha} S_{\alpha})}{\partial t} + \nabla\cdot(\rho_{\alpha}\mathbf{u}_{\alpha})=q_{\alpha},
\end{equation}
where $\phi$ is porosity of porous media, and is assumed to be time-independence, $\alpha=\{w, n\}$ with $w$ and $n$ denoting the wetting and non-wetting phases. $\rho_{\alpha}$, $S_{\alpha}$ and $q_{\alpha}$ are the density, saturation and mass flow rate of phase $\alpha$. $\mathbf{u}_{\alpha}$ is superficial velocity of phase $\alpha$, based on the Darcy's law, it can be given by
\begin{equation}\label{eq2-2}
\mathbf{u}_{\alpha}=-\frac{K_{\alpha}}{\mu_{\alpha}}\nabla P_{\alpha},
\end{equation}
where the gravity effect has been neglected. $\mu_{\alpha}$ and $P_{\alpha}$ are dynamic viscosity and pressure of phase $\alpha$, $K_{\alpha}$ is the effective or apparent permeability, which can can be expressed as
\begin{equation}\label{eq2-3}
K_{\alpha}=K\times k_{r\alpha},
\end{equation}
where $K$ is the absolute permeability of porous media, $k_{r\alpha}$ is the relative permeability of phase $\alpha$, and usually it is a function of saturation $S_{\alpha}$.
If we submit Eqs.~(\ref{eq2-2}) and~(\ref{eq2-3}) into Eq.~(\ref{eq2-1}), the governing equations for saturation $S_{w}$ and $P_{n}$ can be obtained,
\begin{equation}\label{eq2-4}
\phi\frac{\partial (\rho_{w} S_{w})}{\partial t} - \nabla\cdot\big[\rho_{w}\frac{K k_{rw}}{\mu_{w}}\big(\nabla P_{n}-\frac{dP_{c}}{dS_{w}}\nabla S_{w}\big)\big]=q_{w},
\end{equation}
\begin{equation}\label{eq2-5}
\phi\frac{\partial [\rho_{n} (1-S_{w})]}{\partial t} - \nabla\cdot\big(\rho_{n}\frac{K k_{rn}}{\mu_{n}}\nabla P_{n}\big)=q_{n},
\end{equation}
where the following relations have been used to derive above equations,
\begin{equation}\label{eq2-6}
S_{w}+S_{n}=1, \ \ P_{c}=P_{n}-P_{w},
\end{equation}
$P_{c}$ is the capillary pressure, and usually it is also related to saturation $S_{w}$.

It is clear that Eqs.~(\ref{eq2-2}),~(\ref{eq2-4}) and~(\ref{eq2-5}) are strongly coupled \cite{Chen2006}, and the coupling also brings some difficulties in developing efficient and accurate numerical methods. To reduce the coupling, here we introduce the global pressure $P$ and fractional flow function $f_{\alpha}\ (\alpha=w, \ n)$, and after some algebraic manipulations, one can obtain the following global-pressure fractional-flow formulation from Eqs.~(\ref{eq2-2}),~(\ref{eq2-4}) and~(\ref{eq2-5}) \cite{Chen2006,Chavent1986,Cao2011},
\begin{equation}\label{eq2-7}
\mathbf{u}=-K\lambda_{t}\nabla P,\ \ \nabla\cdot\mathbf{u}=\frac{q_{w}}{\rho_{w}}+\frac{q_{n}}{\rho_{n}},
\end{equation}
\begin{equation}\label{eq2-8}
\phi\frac{\partial S_{w}}{\partial t} +\nabla\cdot\big(K\lambda_{n}f_{w}\frac{dP_{c}}{dS_{w}}\nabla S_{w}+f_{w}\mathbf{u}\big)=\frac{q_{w}}{\rho_{w}},
\end{equation}
where $\rho_{w}$ and $\rho_{n}$ have been assumed to be constants, $\mathbf{u}=\mathbf{u}_{w}+\mathbf{u}_{n}$ is the total velocity. $\lambda_{t}=\lambda_{w}+\lambda_{n}$ is the total mobility with $\lambda_{\alpha}\ (\alpha=w, \ n)$ representing the phase mobility,
\begin{equation}\label{eq2-9}
\lambda_{\alpha}=\frac{k_{r\alpha}}{\mu_{\alpha}}.
\end{equation}
The the global pressure $P$ is defined by
\begin{equation}\label{eq2-11}
P=\frac{P_{w}+P_{n}}{2}+\frac{1}{2}\int_{S_{c}}^{S_{w}}\frac{\lambda_{n}-\lambda_{w}}{\lambda_{t}}\frac{dP_{c}}{d\xi}d\xi,
\end{equation}
where $P_{c}(S_{c})=0$. Usually, the relation $S_{c}=1-S_{nr}$ is adopted, and $S_{nr}$ is the residual non-wetting saturation \cite{Chavent1986,Cao2011}.
The fractional flow function $f_{\alpha}\ (\alpha=w, \ n)$ can be expressed as
\begin{equation}\label{eq2-10}
f_{\alpha}=\frac{\lambda_{\alpha}}{\lambda_{t}}.
\end{equation}
 Actually, once the saturation $S_{w}$ and total velocity $\mathbf{u}$ are derived from Eqs.~(\ref{eq2-7}) and~(\ref{eq2-8}), we can also obtain the phase velocities \cite{Chen2006},
\begin{equation}\label{eq2-12}
\mathbf{u}_{w}=f_{w}(\mathbf{u}+K\lambda_{n}\nabla P_{c}),
\end{equation}
\begin{equation}\label{eq2-13}
\mathbf{u}_{n}=f_{n}(\mathbf{u}-K\lambda_{w}\nabla P_{c}).
\end{equation}
From Eqs.~(\ref{eq2-7}) and~(\ref{eq2-8}), it is also obvious that the mathematical model for two-phase flow in porous media consists of one Poisson equation (PE) for pressure and one convection-diffusion equation (CDE) for saturation,
\begin{equation}\label{eq2-15}
\nabla\cdot(D_{p}\nabla P)+F_{p}=0,
\end{equation}
\begin{equation}\label{eq2-14}
\phi\frac{\partial S_{w}}{\partial t} + \nabla\cdot(f_{w}\mathbf{u})= \nabla\cdot(D_{s}\nabla S_{w})+F_{s},
\end{equation}
where the parameters $D_{p}$ and $D_{s}$ may not be constants, and they are given by
\begin{equation}\label{eq2-16}
D_{p}=K\lambda_{t}, \ \ D_{s}=-K\lambda_{n}f_{w}\frac{dP_{c}}{dS_{w}}.
\end{equation}
$F_{s}$ and $F_{p}$ are source terms, and are defined as
\begin{equation}\label{eq2-17}
F_{p}=\frac{q_{w}}{\rho_{w}}+\frac{q_{n}}{\rho_{n}}, \ \ F_{s}=\frac{q_{w}}{\rho_{w}}.
\end{equation}
To eliminate the difficulty of LB method in treating the convection term, we rewrite the CDE~(\ref{eq2-14}) as
\begin{equation}\label{eq2-18}
\phi\frac{\partial S_{w}}{\partial t} = \nabla\cdot(D_{s}\nabla S_{w})+F_{s} + \nabla\cdot(\lambda_{w}K\nabla P),
\end{equation}
where Eq.~(\ref{eq2-7}) has been used.

In the following, the mathematical model composed of Eqs.~(\ref{eq2-15}) and (\ref{eq2-18}) for two-phase flow in porous media would be considered.

\section{Lattice Boltzmann model for two-phase flow in porous media}\label{section3}

In the past decades, many LB models have been developed for Navier-Stokes equations and CDEs. Based on the collision term, however, they can be classified into several categories, i.e., the BGK or single-relaxation-time model \cite{Qian1992,Chen1998,Guo2013,Kruger2017}, the entropic
LB model \cite{Ansumali2002,Ansumali2003}, the two-relaxation-time (TRT) model \cite{Ginzburg2005a,Ginzburg2008}, the multiple-relaxation-time (MRT) model (or the generalized LB model) \cite{dHumieres1992,Lallemand2000}, and central moment model \cite{Geier2006,Premnath2011}. In this work, although we only focus on the BGK model for its simplicity and computational efficiency, there are no substantial difficulties to extend present BGK model to a more general MRT model.

In the framework of LB method, PE is usually treated as the steady diffusion equation, and thus the LB models for diffusion equation can be directly used for the PE \cite{He2000,Hirabayashi2001,Wang2006,Wang2011}. It should be noted that, however, some undesirable errors may be induced by the inappropriate initialization when these LB models for diffusion equation are used to solve the PE \cite{Chai2008}. To eliminate the undesirable errors caused by initialization, Chai and Shi proposed a genuine LB model for PE \cite{Chai2008}. Based the this work, the LB equation of present model for Eq.~(\ref{eq2-15}) can be written as \cite{Chai2008}
\begin{equation}\label{eq3-1}
f_{i}(\mathbf{x} + \mathbf{c}_{i}\delta t, t' + \delta t) =  f_{i}(\mathbf{x}, t' )-\frac{1}{\tau_{f}}\big[f_{i}(\mathbf{x}, t') - f_{i}^{(eq)}(\mathbf{x}, t')\big]+\delta t\bar{\omega}_{i}F_{p},
\end{equation}
where $f_{i}(\mathbf{x}, t')$ is the distribution function associated with the discrete velocity $\mathbf{c}_{i}$ at position $\mathbf{x}$ and pseudo time $t'$. It should be noted that the pseudo time $t'$ is independent of physical time $t$ appeared in section 2 \cite{Meng2016}. In our simulations,
Eq.~(\ref{eq3-1}) is iterated to reach a steady state such that the pressure $P$ satisfying Eq.~(\ref{eq2-15}) can be
obtained at the physical time $t$. $f_{i}^{(eq)}(\mathbf{x}, t')$ is the equilibrium distribution function, and can be defined as
\begin{equation}\label{eq3-3}
f_{i}^{(eq)}(\mathbf{x}, t')=\left\{\begin{array}{c} (\omega_{0}-1)P,\ \ \   \emph{i}=0\\
\omega_{i}P, \ \ \ \ \ \ \ \ \ \ \ \emph{i}\neq 0
\end{array}\right.
\end{equation}
where $\omega_{i}$ and $\bar{\omega}_{i}$ are the weight coefficients.

On the other hand, there are also some LB models for the diffusion equations or CDEs \cite{Wolf-Gladrow1995,Dawson1993,Ginzburg2005a,Shi2009,Chopard2009,Huber2010,Yoshida2010,Du2013,Huang2014,Chai2013,Chai2014,Yang2014,Chai2016,Aursjo2017,Li2017}. Based on the recent works \cite{Chai2013,Chai2014,Yang2014,Chai2016,Aursjo2017}, the LB equation of present model for Eq.~(\ref{eq2-18}) reads
\begin{eqnarray}\label{eq3-2}
g_{i}(\mathbf{x} + \mathbf{c}_{i}\delta t, t + \delta t) & = & g_{i}(\mathbf{x}, t )-\frac{1}{\tau_{g}}\big[g_{i}(\mathbf{x}, t) - g_{i}^{(eq)}(\mathbf{x}, t)\big]+\delta t\omega_{i}\big(1-\frac{1}{2\tau_{g}}\big)F_{s}\nonumber \\
& + & \gamma\delta t\omega_{i}\frac{\mathbf{c}_{i}\cdot\nabla P}{\tau_{g}},
\end{eqnarray}
where $g_{i}(\mathbf{x},\;t)$ is the distribution function at position $\mathbf{x}$ and time $t$, $\gamma$ is a parameter, and to be determined later. $g_{i}^{(eq)}(\mathbf{x},\;t)$ is the equilibrium distribution function, and can be given by \cite{Shi2009,Chai2016}
\begin{equation}\label{eq3-4}
g_{i}^{(eq)}(\mathbf{x}, t)=\omega_{i}S_{w}\big[\phi+\frac{\mathbf{C}:(\mathbf{c}_{i}\mathbf{c}_{i}-c_{s}^{2}\mathbf{I})}{2c_{s}^{2}}\big], \ \ \mathbf{C}=(\beta-\phi)\mathbf{I},
\end{equation}
where $\mathbf{I}$ is the unit matrix, $c_{s}$ is a parameter related to lattice speed $c=\delta x/\delta t$, $\delta x$ and $\delta t$ are the lattice spacing and time step, respectively. $\beta$ is a parameter that can be used to adjust the relaxation parameter $\tau_{g}$ for a fixed diffusion coefficient $D_{s}$, and also, to ensure that the model is stable, the parameter $\beta$ should satisfy the relation $\phi/2\leq\beta\leq2\phi$ such that the $g_{i}^{(eq)}$ can be non-negative.
The last term in the right hand of Eq.~(\ref{eq3-2}) can also be viewed as a source term, and to correctly recover Eq.~(\ref{eq2-18}) from Eq.~(\ref{eq3-2}), the parameter $\gamma$ should be determined by
\begin{equation}\label{eq3-8}
\gamma=-\frac{\lambda_{w}K}{c_{s}^{2}\delta t}.
\end{equation}

We would like to point out that above LB model can be applied for one, two, and three-dimensional two-phase flows in porous media. Here we only take the two-dimensional case as an example, and consider the D2Q9 lattice model (nine discrete directions in two-dimensional space) \cite{Qian1992} where the weight coefficient, discrete velocity, and relation between parameter $c_{s}$ and lattice speed $c$ can be expressed as
\begin{equation}\label{eq3-5}
\omega_{0}=\frac{4}{9},\ \ \omega_{i=1-4}=\frac{1}{9}, \ \ \omega_{i=5-8}=\frac{1}{36},
\end{equation}
\begin{equation}\label{eq3-5a}
\bar{\omega}_{0}=0,\ \ \bar{\omega}_{i=1-4}=\frac{1}{8}, \ \ \bar{\omega}_{i=5-8}=\frac{1}{8},
\end{equation}
\begin{equation}\label{eq3-6}
\mathbf{c}_{i}=\left\{\begin{array}{ll} (0,\ 0),\ \  \ \ \ \ \ \ \  \ \ \ \ \ \ \ \ \ \ \ \ \ \ \ \ \ \ \ \ \ \ \ \ \ \ \ \ \ \  \ \ \ \ \ \ \ \ \emph{i}=0\\
(\cos[(i-1)\pi/2],\ \sin[(i-1)\pi/2])c, \ \ \ \ \ \ \ \ \ \ \ \emph{i}=1-4\\
(\cos[(2i-9)\pi/4],\ \sin[(2i-9)\pi/4])\sqrt{2}c, \ \ \ \  \emph{i}=5-8\\
\end{array}\right.
\end{equation}
\begin{equation}\label{eq3-6a}
c_{s}^{2}=\frac{1}{3}c^{2}.
\end{equation}
From Eqs.~(\ref{eq3-5}), (\ref{eq3-6}) and (\ref{eq3-6a}), it can be shown that the equilibrium distribution functions $f_{i}^{(eq)}$ and $g_{i}^{(eq)}$ satisfy the following conditions,
\begin{subequations}\label{eq3-7}
\begin{equation}
\sum_{i=0}^{8}f_{i}^{(eq)}=0, \ \ \sum_{i=0}^{8}\mathbf{c}_{i}f_{i}^{(eq)}=\mathbf{0}, \ \ \sum_{i=0}^{8}\mathbf{c}_{i}\mathbf{c}_{i}f_{i}^{(eq)}=Pc_{s}^{2}\mathbf{I},
\end{equation}
\begin{equation}
\sum_{i=0}^{8}g_{i}^{(eq)}=\phi S_{w}, \ \ \sum_{i=0}^{8}\mathbf{c}_{i}g_{i}^{(eq)}=\mathbf{0}, \ \ \sum_{i=0}^{8}\mathbf{c}_{i}\mathbf{c}_{i}g_{i}^{(eq)}=\beta S_{w}c_{s}^{2}\mathbf{I}.
\end{equation}
\end{subequations}

The pressure $P$ and saturation $S_{w}$ are computed by
\begin{subequations}\label{eq3-9}
\begin{equation}
P=\frac{1}{1-\omega_{0}}\sum_{i=1}^{8}f_{i},
\end{equation}
\begin{equation}
S_{w}=\frac{1}{\phi}\big(\sum_{i=0}^{8}g_{i}+\frac{1}{2}\delta t F_{s}\big),
\end{equation}
\end{subequations}
and simultaneously, the pressure gradient and the velocity can also be calculated through the following equations (see \textbf{Appendix A} for details),
\begin{subequations}\label{eq3-10}
\begin{equation}
\nabla P =  -\frac{\sum_{i}\mathbf{c}_{i}[f_{i}-f_{i}^{(eq)}]}{\tau_{f}\delta t c_{s}^{2}} =  -\frac{\sum_{i}\mathbf{c}_{i}f_{i}}{\tau_{f}\delta t c_{s}^{2}},
\end{equation}
\begin{equation}
\mathbf{u}=-K\lambda_{t}\nabla P = \frac{K\lambda_{t}\sum_{i}\mathbf{c}_{i}f_{i}}{\tau_{f}\delta t c_{s}^{2}},
\end{equation}
\end{subequations}
where Eq.~(\ref{eq3-7}a) has been used to derive Eq.~(\ref{eq3-10}a).

Through the Chapman-Enskog analysis, one can find that the nonlinear PE (\ref{eq2-15}) and CDE (\ref{eq2-18}) can be recovered correctly from the present LB model, and the parameters $D_{p}$ and $D_{s}$ are related to relaxation time $\tau_{f}$ and $\tau_{g}$,
\begin{subequations}\label{eq3-11}
\begin{equation}
D_{p}=c_{s}^{2}\big(\tau_{f}-\frac{1}{2}\big)\delta t,
\end{equation}
\begin{equation}
D_{s}=\beta c_{s}^{2}\big(\tau_{g}-\frac{1}{2}\big)\delta t.
\end{equation}
\end{subequations}
Finally, some remarks on the present LB model are listed as follows.

\emph{Remark I}: We would like to point that although Eq.~(\ref{eq2-14}) or (\ref{eq2-18}) is similar to the commonly used CDE, there is a great difference due to the appearance of porosity $\phi$. Actually, if $\phi$ is a constant, we can rewrite Eq.~(\ref{eq2-14}) or (\ref{eq2-18}) in standard form through dividing $\phi$ on the both sides of Eq.~(\ref{eq2-14}) or (\ref{eq2-18}), then some available LB models for standard CDEs \cite{Dawson1993,Ginzburg2005a,Chopard2009,Huber2010,Yoshida2010,Du2013,Huang2014,Yang2014,Chai2016,Li2017} can be applied. If $\phi$ is space-dependence rather than a constant, however, there would be some difficulties in rewriting Eq.~(\ref{eq2-14}) or (\ref{eq2-18}) as a classical CDE, and these available LB models for CDEs \cite{Dawson1993,Chopard2009,Huber2010,Yoshida2010,Du2013,Huang2014,Yang2014,Chai2016,Li2017} cannot be directly used to solve Eq.~(\ref{eq2-14}) or (\ref{eq2-18}). We note that Ginzburg has proposed two LB models (E and L models) for general CDEs \cite{Ginzburg2005a}, and these two models also seem to be suitable for Eq.~(\ref{eq2-14}). However, as pointed out by in Ref. \cite{Shi2009}, some additional assumptions on the convection term have been adopted to recover correct CDEs. On the contrary, in the present LB equation for Eq.~(\ref{eq2-14}) or (\ref{eq2-18}), the convection term is considered as a source term, and thus the difficulty of LB method in treating the convection term and/or some assumptions on the convection term \cite{Dawson1993,Ginzburg2005a,Chopard2009,Huber2010,Du2013,Yang2014} can be eliminated.

\emph{Remark II}: As stated previously, the convection term is treated as a source term. To include effect of the source term, we added a term related to space derivative in the LB equation for saturation [see Eq.~(\ref{eq3-2})], which also leads to the fact that the collision process cannot be implemented locally when nonlocal finite-difference schemes are used to calculate the space-derivative term. However, in the present work, the collision process can be conducted locally since the space-derivative term can be computed locally from Eq.~(\ref{eq3-10}a).

\emph{Remark III}: Similar to the results in Refs. \cite{Chai2013,Chai2014,Yang2014,Meng2016}, in the present LB model, the pressure gradient and velocity $\mathbf{u}$ can also be computed locally by Eq.~(\ref{eq3-10}), and also have a second-order convergence rate in space (see the results in the following section).

\emph{Remark IV}: In the global-pressure fractional-flow model for two-phase flow in porous media, the parameters $D_{p}$ and $D_{s}$ are usually function of $S_{w}$ rather than constants, which would also cause the relaxation time $\tau_{f}$ and $\tau_{g}$ in the present LB model to be space-dependence. However, if $D_{p}$ is a constant, the LB model in Ref. \cite{Chai2008} can also be applied for PE (\ref{eq2-15}).

\emph{Remark V}: We would like to point out that to reduce the computational cost, one can also consider the simple D2Q4 and D2Q5 lattice models \cite{Cui2016} for two-dimensional problems, but the equilibrium distribution function (\ref{eq3-4}) should be modified to satisfy the condition (\ref{eq3-7}b). Actually, in the DnQ(2n) and DnQ(2n+1) lattice models for n-dimensional problems, the following equilibrium distribution function can be adopted,
\begin{equation}\label{eq3-4b}
g_{i}^{(eq)}(\mathbf{x}, t)=\left\{\begin{array}{ll} \big[(\omega_{0}-1)\beta +\phi\big]S_{w},\ \ \ \ \ \ \ \  \ \emph{i}=0\\
\omega_{i}\beta S_{w}, \ \ \ \ \ \ \ \ \ \ \ \ \ \ \ \ \ \ \ \ \ \ \ \ \emph{i}\neq 0
\end{array}\right.
\end{equation}
where the parameter $\beta$ should satisfy the relation $0\leq\beta\leq\phi/(1-\omega_{0})$ to make $g_{i}^{(eq)}$ non-negative.

\section{Numerical results and discussion}

In this section, we would perform some simulations to validate present LB model for two-phase flows in porous media. Unless otherwise stated, the parameter $\beta$ is set to be 1.0, the anti-bounce-back scheme is applied for Dirichlet boundary condition \cite{Yoshida2010,Ginzburg2005b,Zhang2012,Li2013,Yong2015,Yong2017,Cui2016},
\begin{equation}
   f_{i}(\mathbf{x}_{f}, t'+\delta t)=-f_{\bar{i}}^{+}(\mathbf{x}_{f}, t')+2\omega_{\bar{i}}P_{b}, \ \ g_{i}(\mathbf{x}_{f}, t+\delta t)=-g_{\bar{i}}^{+}(\mathbf{x}_{f}, t)+2g_{\bar{i}}^{(eq)}(S_{wb}),
\end{equation}
and the classical bounce-back scheme is adopted for no-flux boundary condition \cite{Ginzburg2005b,Chai2016b},
\begin{equation}
   f_{i}(\mathbf{x}_{f}, t'+\delta t)=f_{\bar{i}}^{+}(\mathbf{x}_{f}, t'), \ \ g_{i}(\mathbf{x}_{f}, t+\delta t)=g_{\bar{i}}^{+}(\mathbf{x}_{f}, t),
\end{equation}
where $P_{b}$ and $S_{wb}$ are specified values of pressure and saturation at boundaries, $\bar{i}$ is the opposite direction of $i$ (e.g., $i=1$, $\bar{i}=3$). $f_{i}(\mathbf{x}_{f}, t'+\delta t')$ and $g_{i}(\mathbf{x}_{f}, t+\delta t)$ are the unknown distribution functions at the node $\mathbf{x}_{f}$, $f_{\bar{i}}^{+}(\mathbf{x}_{f}, t')$ and $g_{\bar{i}}^{+}(\mathbf{x}_{f}, t)$ are post-collision distribution functions, and are given by
\begin{subequations}\label{2-1b}
\begin{equation}
f_{i}^{+}(\mathbf{x}, t') = f_{i}(\mathbf{x}, t' )-\frac{1}{\tau_{f}}\big[f_{i}(\mathbf{x}, t') - f_{i}^{(eq)}(\mathbf{x}, t')\big]+\delta t\bar{\omega}_{i}F_{p},
\end{equation}
\begin{equation}
g_{i}^{+}(\mathbf{x}, t) = g_{i}(\mathbf{x}, t )-\frac{1}{\tau_{g}}\big[g_{i}(\mathbf{x}, t) - g_{i}^{(eq)}(\mathbf{x}, t)\big]+\delta t\omega_{i}\big(1-\frac{1}{2\tau_{g}}\big)F_{s}+\gamma\delta t\omega_{i}\frac{\mathbf{c}_{i}\cdot\nabla P}{\tau_{g}}.
\end{equation}
\end{subequations}

In the initialization process, the distribution function $g_{i}$ for saturation can be approximately given by
\begin{equation}\label{gini}
g_i(\mathbf{x}, t=0)\approx g_i^{(eq)}(\mathbf{x}, t=0)+\epsilon g_i^{(1)}(\mathbf{x}, t=0).
\end{equation}
Actually, the initial value of equilibrium distribution function $g_i^{(eq)}(\mathbf{x}, t=0)$ can be directly obtained through the initial condition of saturation $S_{w}(\mathbf{x}, t=0)$, while the non-equilibrium part $\epsilon g_i^{(1)}(\mathbf{x}, t=0)$ is unknown, and needs to be determined. Based on the Chapman-Enskog analysis [see Eq.~(\ref{ConB2})], the non-equilibrium part $\epsilon g_i^{(1)}(\mathbf{x}, t=0)$ can be expressed by
\begin{eqnarray}\label{gneq}
\epsilon g_{i}^{(1)}(\mathbf{x}, t=0) & = & -\tau_{g}\delta t\big[\epsilon\bar{D}_{i1}g_{i}^{(0)}-\omega_{i}\big(1-\frac{1}{2\tau_{g}}\big) F_{s}-\gamma\omega_{i}\frac{\mathbf{c}_{i}\cdot\nabla P}{\tau_{g}}\big]|_{t=0}\nonumber\\
& = & -\tau_{g}\delta t\big[\Gamma_{i}^{(0)}\epsilon\bar{D}_{i1}S_{w}-\omega_{i}\big(1-\frac{1}{2\tau_{g}}\big) F_{s}-\gamma\omega_{i}\frac{\mathbf{c}_{i}\cdot\nabla P}{\tau_{g}}\big]|_{t=0}\nonumber\\
& = & -\tau_{g}\delta t\big[\Gamma_{i}^{(0)}\big(\frac{F_{s}}{\phi}+\mathbf{c}_{i}\cdot\nabla S_{w}\big)-\omega_{i}\big(1-\frac{1}{2\tau_{g}}\big) F_{s}-\gamma\omega_{i}\frac{\mathbf{c}_{i}\cdot\nabla P}{\tau_{g}}\big]|_{t=0},\nonumber\\
& = & -\tau_{g}\delta t\big[\big(\frac{\Gamma_{i}^{(0)}}{\phi}-\omega_{i}\big(1-\frac{1}{2\tau_{g}}\big)\big)F_{s}+\Gamma_{i}^{(0)}\mathbf{c}_{i}\cdot\nabla S_{w}-\gamma\omega_{i}\frac{\mathbf{c}_{i}\cdot\nabla P}{\tau_{g}}\big]|_{t=0},
\end{eqnarray}
where Eqs.~(\ref{CEB1}) and (\ref{ReB1}) have been used. The function $\Gamma_{i}^{0}$ is defined by
\begin{equation}
\Gamma_{i}^{(0)}=\frac{1}{S_{w}}g_{i}^{(eq)}=\omega_{i}\big[\phi+\frac{\mathbf{C}:(\mathbf{c}_{i}\mathbf{c}_{i}-c_{s}^{2}\mathbf{I})}{2c_{s}^{2}}\big].
\end{equation}
Once the initial value of $S_{w}$ is given, we can also obtain its gradient $\nabla S_{w}$. Then substituting Eq.~(\ref{gneq}) into Eq.~(\ref{gini}), one can derive the initial value of distribution function $g_{i}(\mathbf{x}, t=0)$.

In addition, to test accuracy and convergence rate of the LB model for pressure, pressure gradient or equivalently velocity, and saturation, the following relative error is used,
\begin{equation}
E_{\psi}=\frac{\sum_{\mathbf{x}}|\psi_{a}(\mathbf{x}, t)-\psi_{n}(\mathbf{x}, t)|}{\sum_{\mathbf{x}}|\psi_{a}(\mathbf{x}, t)|},
\end{equation}
where $\psi$ denotes the pressure $P$, one component of pressure gradient or velocity, or saturation $S_{w}$, the subscripts $a$ and $n$ represent its analytical and numerical solutions.

\subsection{Example 1: A simple decoupled problem}

For simplicity, we first considered a simple problem where Eq.~(\ref{eq2-15}) for pressure and Eq.~(\ref{eq2-18}) for saturation are decoupled through setting  $\lambda_{t}=\lambda_{w}=\lambda$, which is the same as the problem in Ref. \cite{Chen1999}. For this special case, we can also rewrite Eq.~(\ref{eq2-18}) as
\begin{equation}\label{eq4-1}
\phi\frac{\partial S_{w}}{\partial t} = \nabla\cdot(D_{s}\nabla S_{w})+Q,
\end{equation}
where $Q=F_{s}-F_{p}$.

The domain of the problem is $\Omega=[0, 2]\times[0, 2]$, $\phi=1.0$, the parameters $D_{p}$ and $D_{s}$ are chosen as two constants, and set to be 0.001. The analytical solutions of pressure, saturation and velocity can be given by
\begin{subequations}\label{eq4-2}
\begin{equation}
P(\mathbf{x}, t)=1.0+\sin{(\pi x)}\sin{(\pi y)},
\end{equation}
\begin{equation}
\mathbf{u}=(u_{x},\ u_{y})^{\intercal}=-D_{p} \nabla P=-D_{p}\pi\big[\cos{(\pi x)}\sin{(\pi y)}, \;\sin{(\pi x)}\cos{(\pi y)}\big]^{\intercal},
\end{equation}
\begin{equation}
S_{w}(\mathbf{x}, t)=t\sin{(\pi x)}\sin{(\pi y)},
\end{equation}
\end{subequations}
where $\intercal$ represents the transpose of a matrix. If we substitute analytical solutions [Eqs.~(\ref{eq4-2}a) and (\ref{eq4-2}c)] into Eqs.~(\ref{eq2-15}) and Eq.~(\ref{eq4-1}), one can determine the source terms $F_{p}$ and $Q$,
\begin{subequations}
\begin{equation}
F_{p}=2D_{p}\pi^{2}\sin{(\pi x)}\sin{(\pi y)},
\end{equation}
\begin{equation}
Q=(2D_{s}\pi^{2}t+\phi)\sin{(\pi x)}\sin{(\pi y)}.
\end{equation}
\end{subequations}

In our simulations, the initial and boundary conditions of the pressure and saturation are given by their analytical solutions, i.e., Eqs. (\ref{eq4-2}a) and (\ref{eq4-2}c). We first performed a simulation with a lattice size $64\times64$, and presented the results of pressure, velocity and saturation at a specified time $T=1.0$ in Figs. 1-4 where $c=1.0$. As seen from these figures, the numerical results are in good agreement with the corresponding analytical solutions, and the global relative errors of pressure, velocity and saturation are less than $3.27\times10^{-3}$.

\begin{figure}
\includegraphics[width=2.5in]{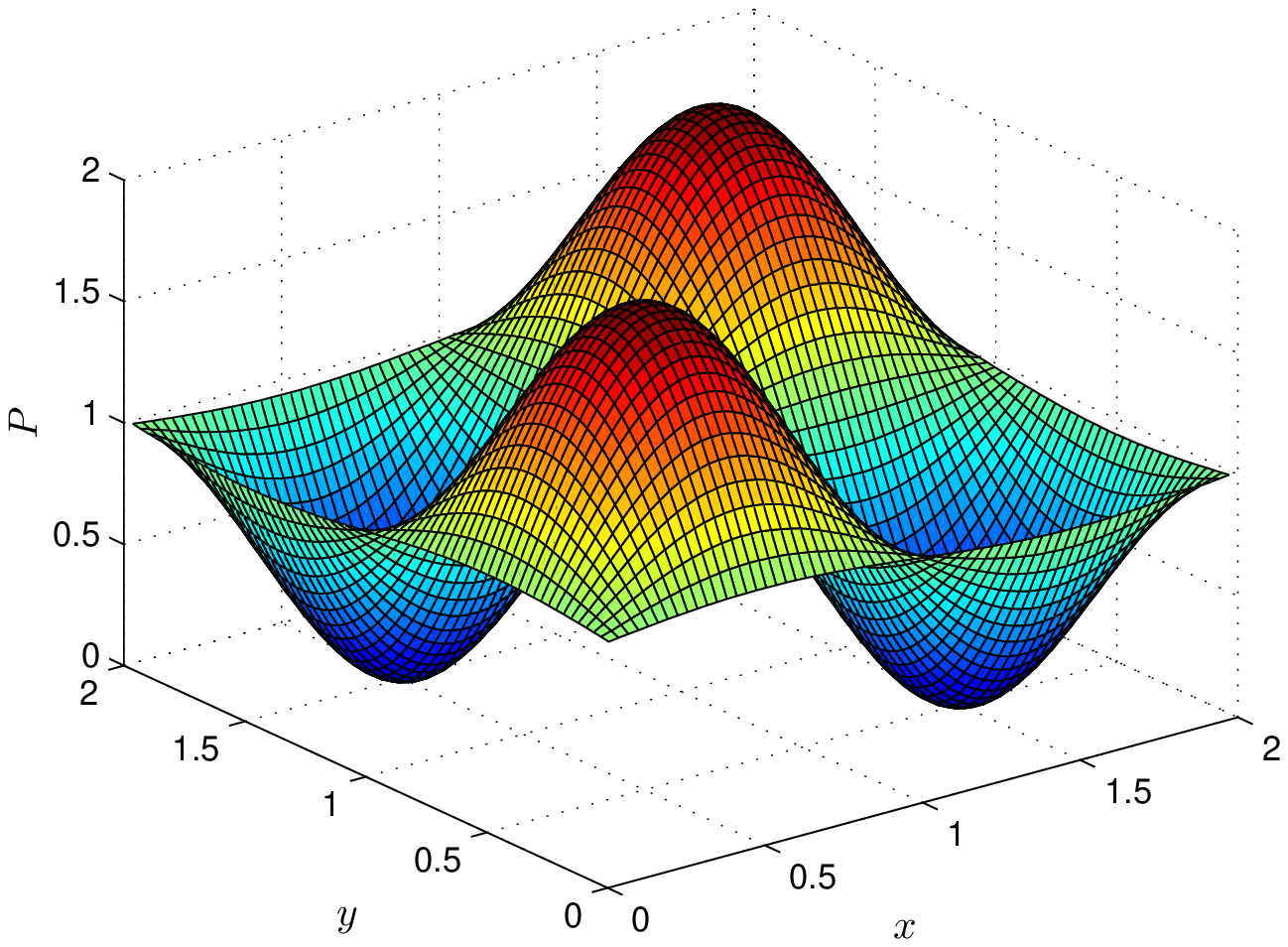}
\includegraphics[width=2.5in]{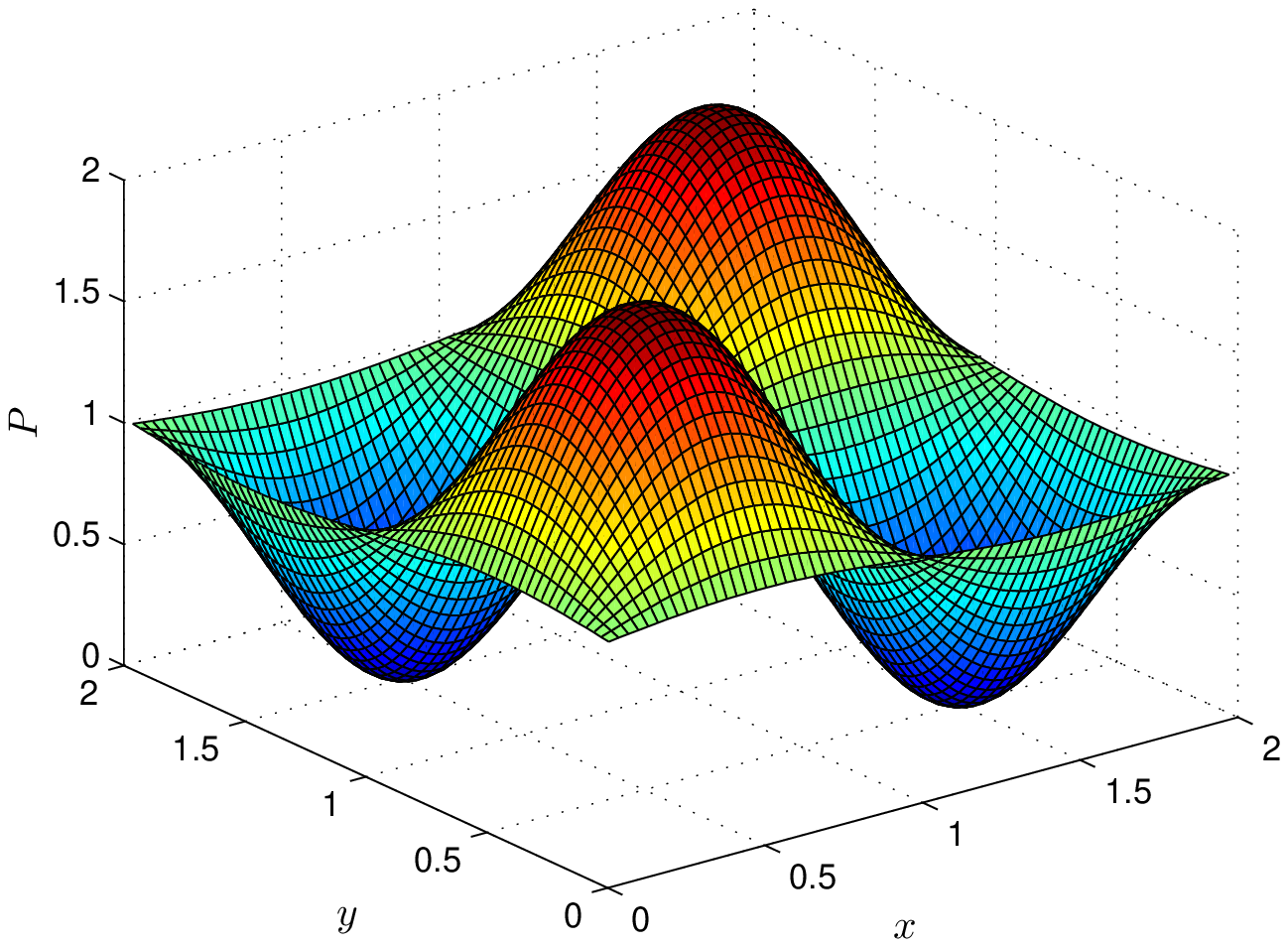}
\centering\caption{\label{fig:1} Distributions of pressure $P$ [(a): analytical solution, (b): numerical solution].}
\end{figure}

\begin{figure}
\includegraphics[width=2.5in]{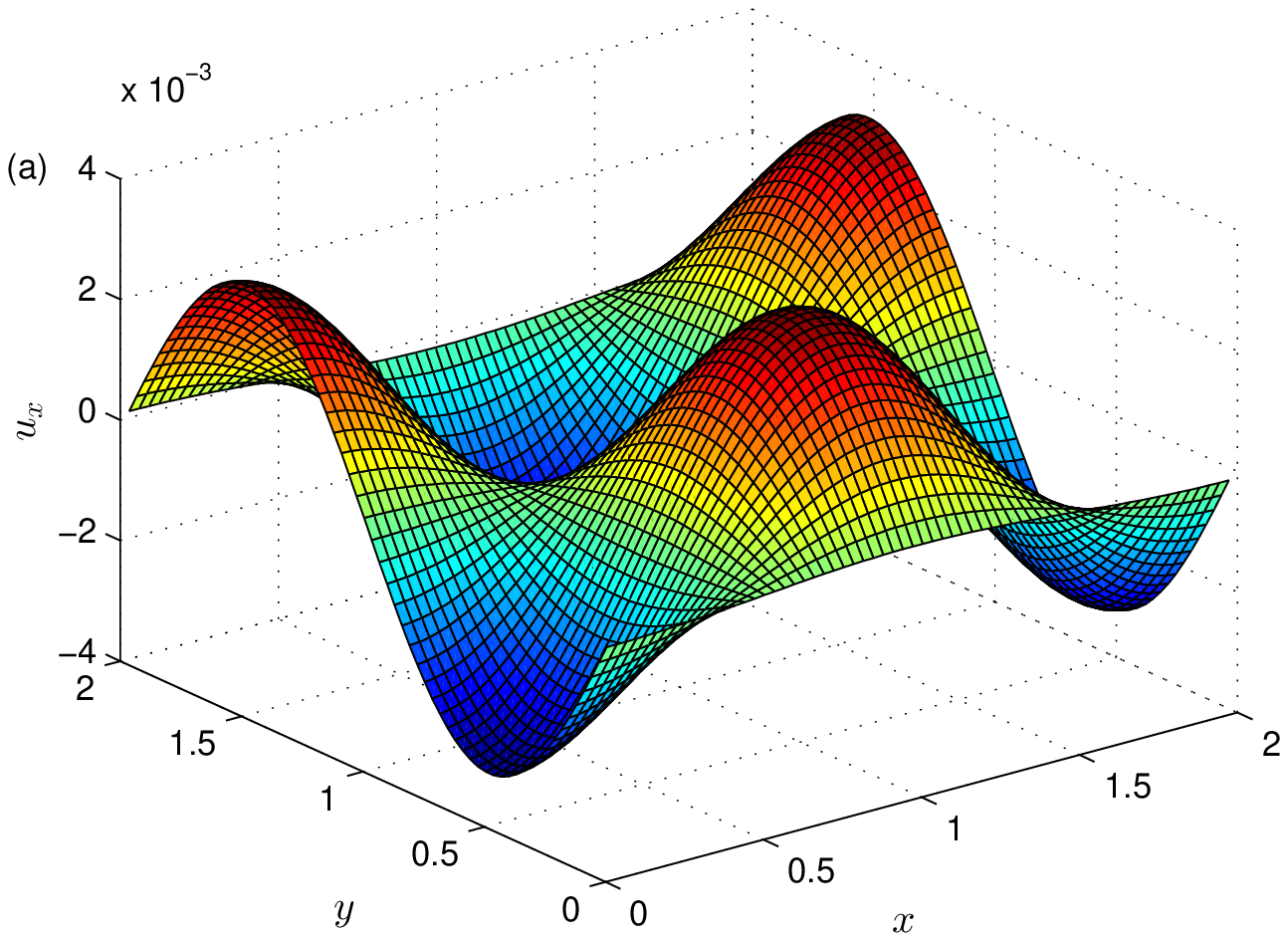}
\includegraphics[width=2.5in]{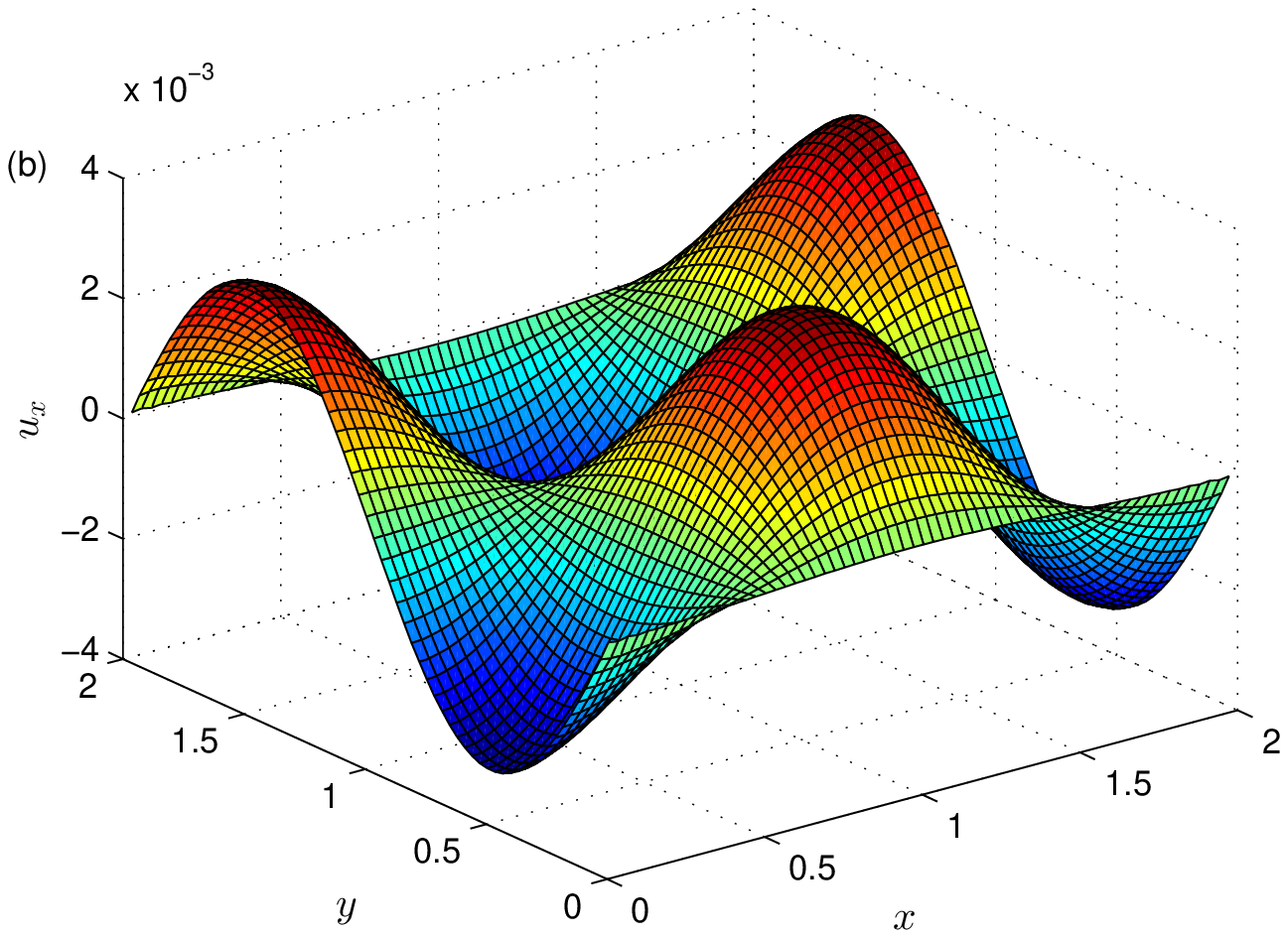}
\centering\caption{\label{fig:2} Distributions of velocity component $u_{x}$ [(a): analytical solution, (b): numerical solution].}
\end{figure}

\begin{figure}
\includegraphics[width=2.5in]{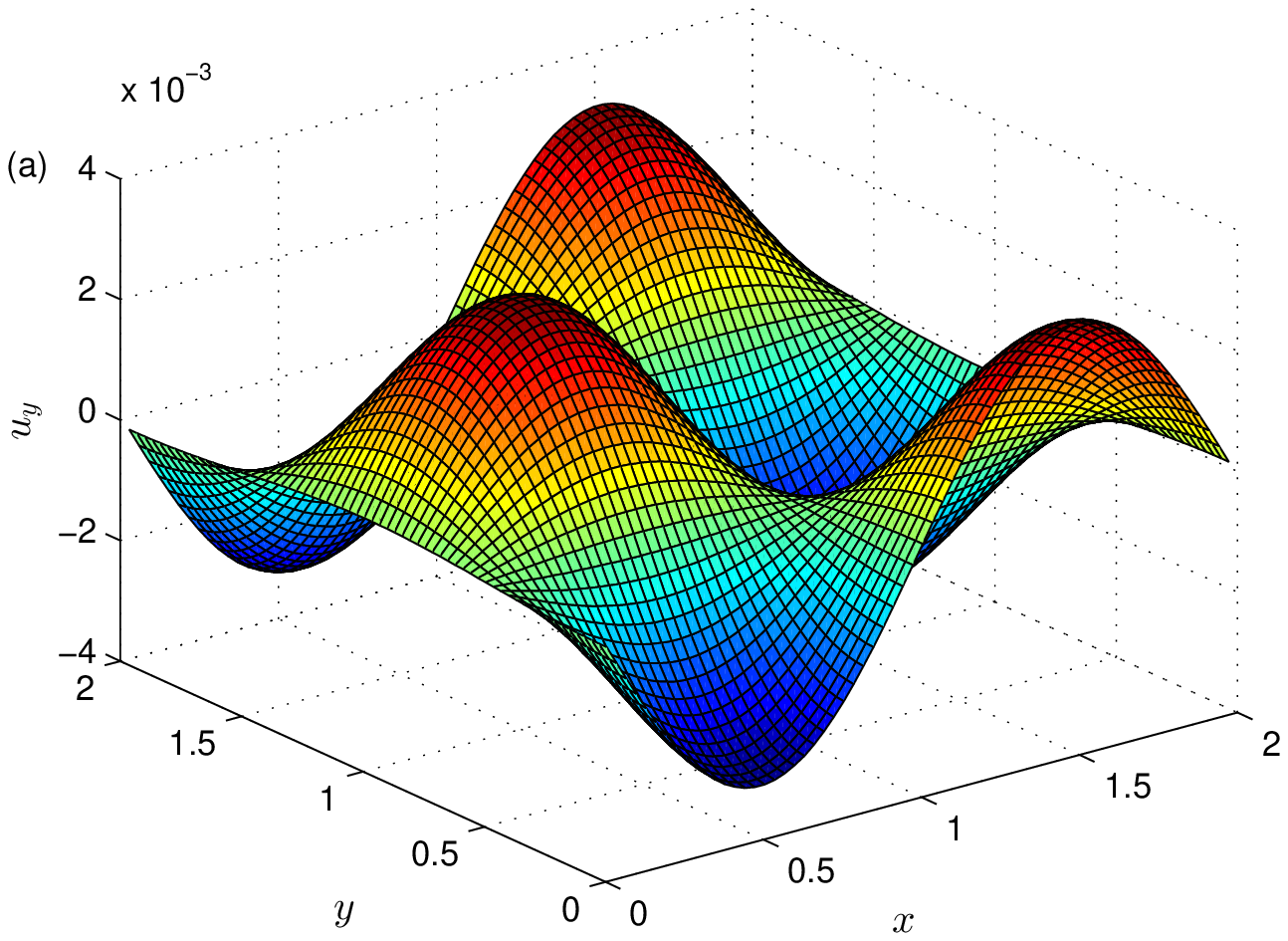}
\includegraphics[width=2.5in]{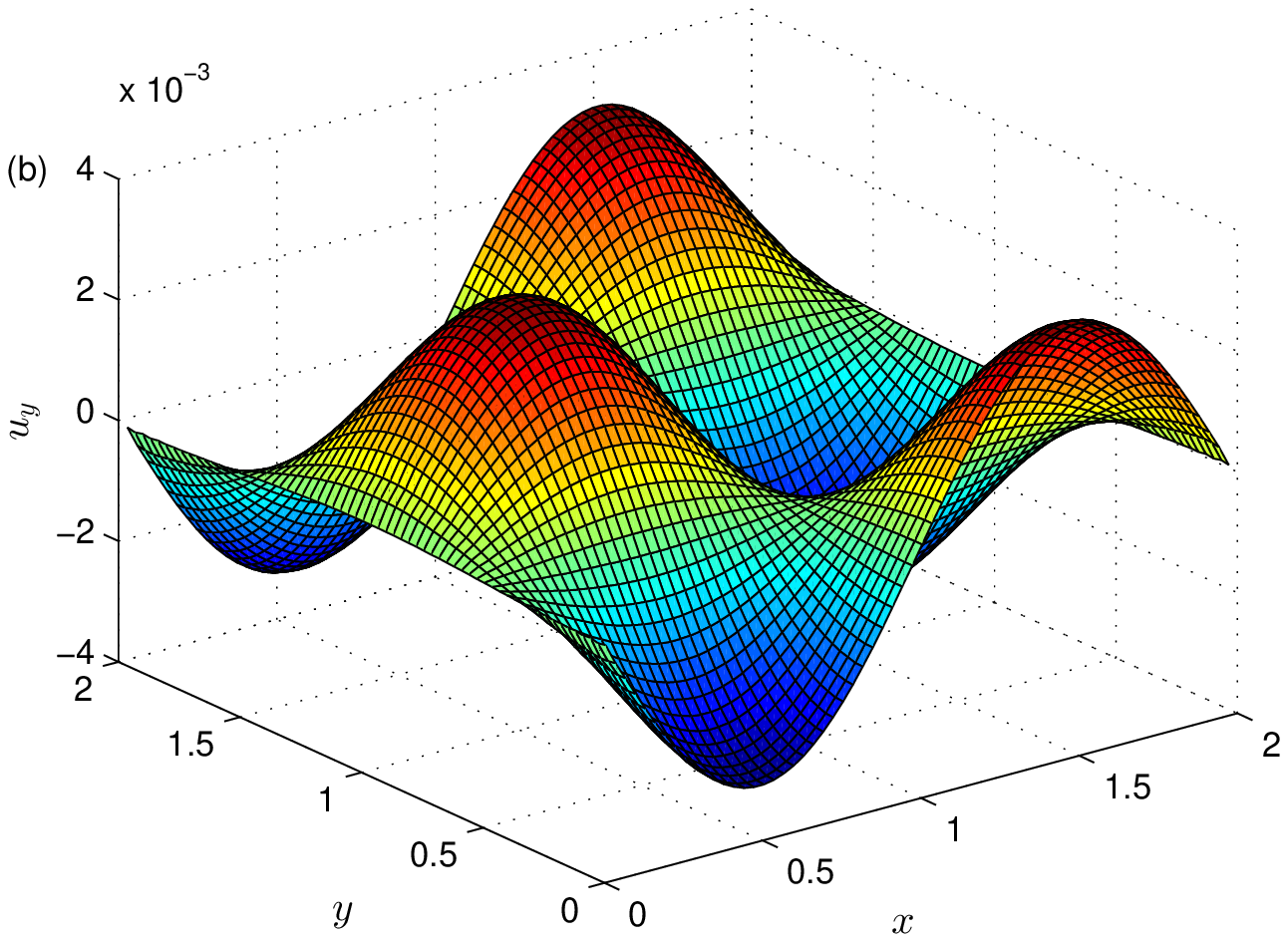}
\centering\caption{\label{fig:3} Distributions of velocity component $u_{y}$ [(a): analytical solution, (b): numerical solution].}
\end{figure}

\begin{figure}
\includegraphics[width=2.5in]{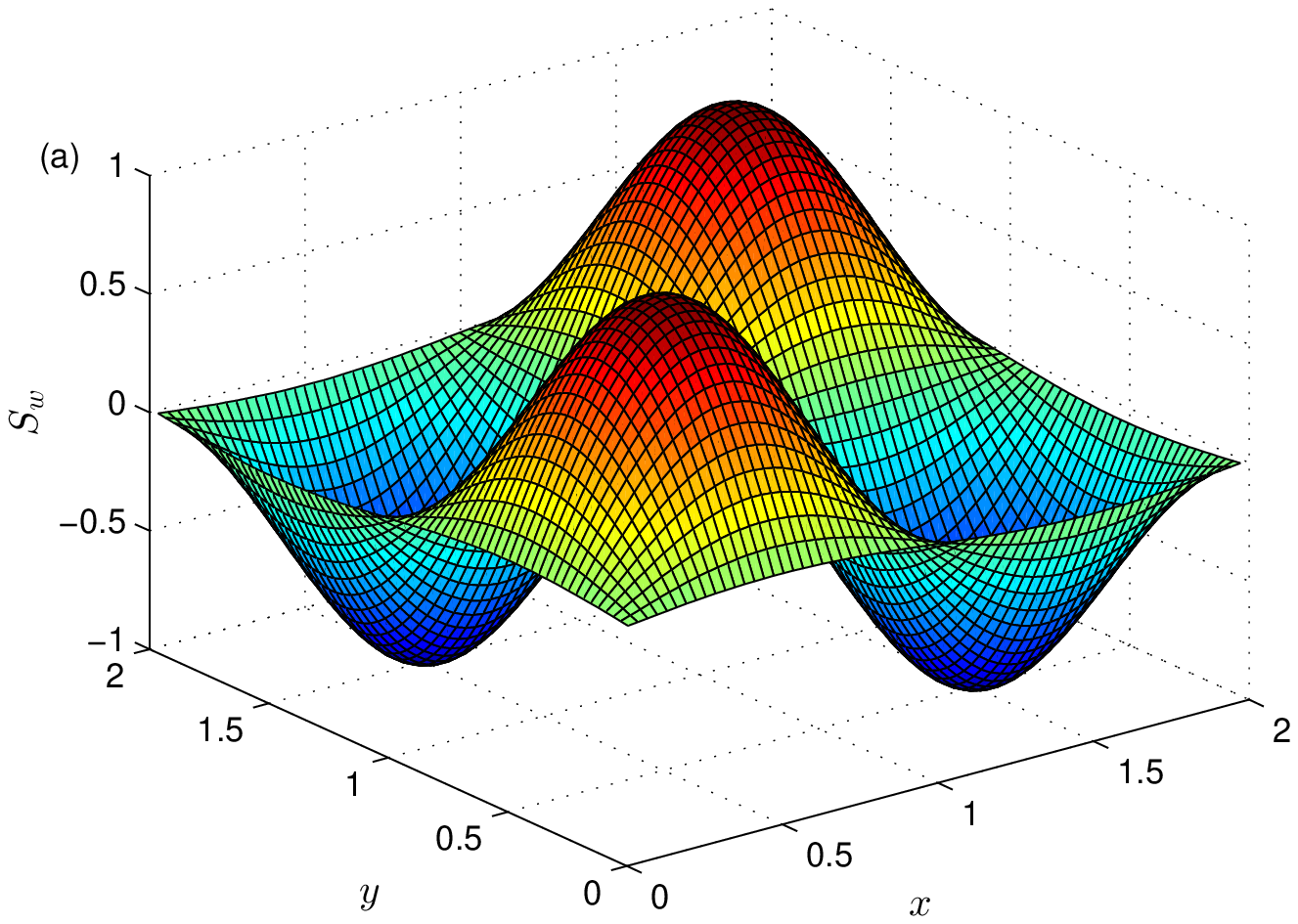}
\includegraphics[width=2.5in]{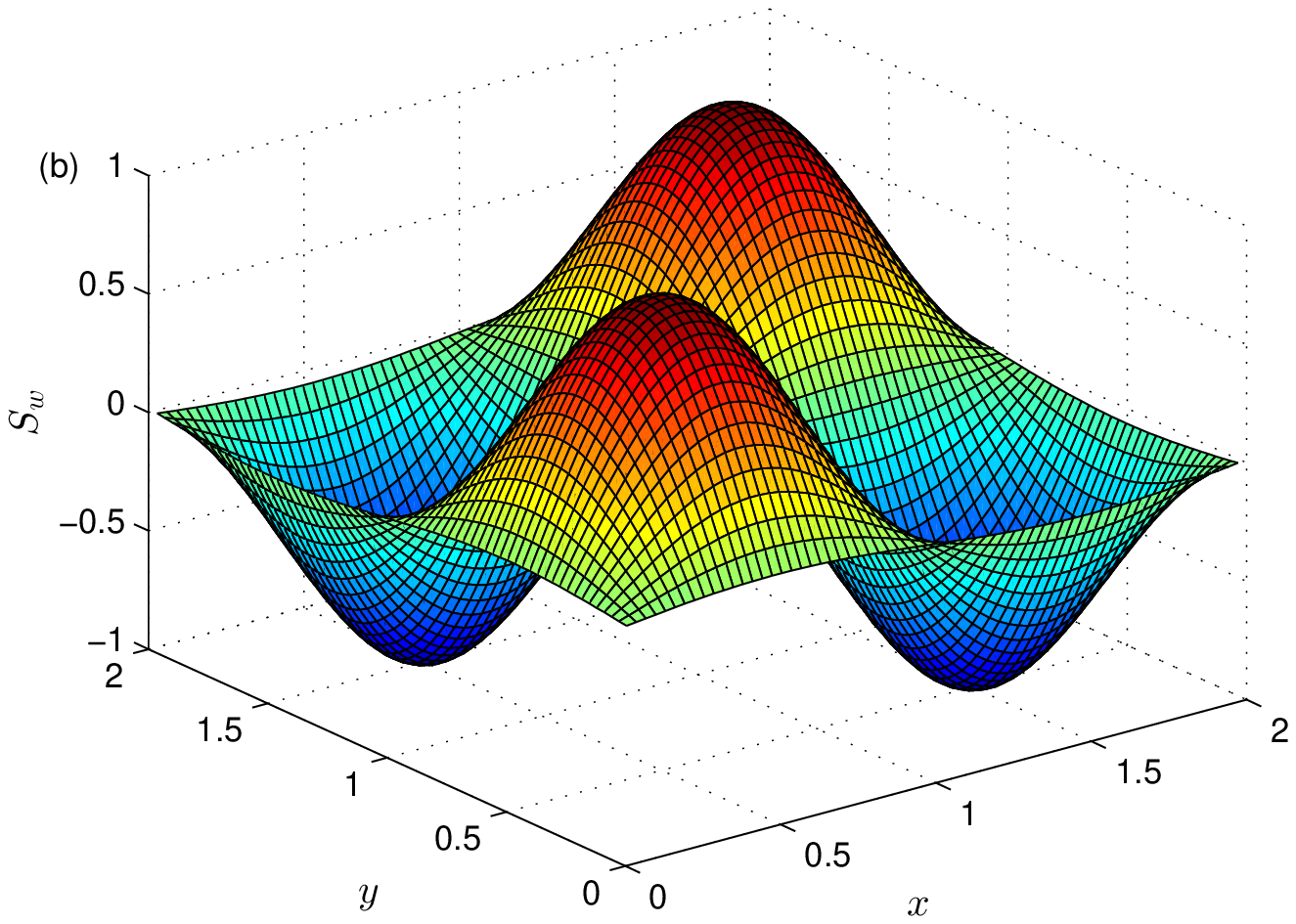}
\centering\caption{\label{fig:4} Distributions of saturation $S_{w}$ [(a): analytical solution, (b): numerical solution].}
\end{figure}

Then the problem is also used to test the convergence rate of present LB model since it is very simple, and the mathematical equations for pressure and saturation are decoupled. We carried out several simulations with different lattice sizes, and calculated the global relative errors in Fig. 5. As shown in this figure, the present LB model has a second-order convergence rate in computing pressure, velocity, and saturation.

\begin{figure}
\includegraphics[width=3in]{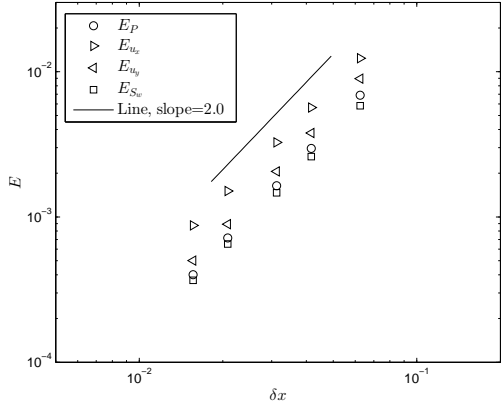}
\centering\caption{\label{fig:4} The global relative errors of pressure, velocity and saturation at different lattice sizes ($\delta x=L/128, \ L/96,\ L/64,\ L/48,\ L/32$, $L=2.0$). The slope of inserted line is 2.0, which indicates that the present LB model has a second-order convergence rate.}
\end{figure}

\subsection{Example 2: A coupled benchmark problem}

In this part, we continue to consider another benchmark problem where the governing equations for pressure and saturation, i.e., Eqs.~(\ref{eq2-15}) and (\ref{eq2-18}), are nonlinearly coupled. Following the previous works \cite{Ohlberger1997,Mozolevski2013}, some parameters appeared in Eqs.~(\ref{eq2-15}) and (\ref{eq2-18}) are given by
\begin{subequations}\label{eq4-4}
\begin{equation}
D_{p}=\frac{1}{0.5-0.2S_{w}}, \ \ D_{s}=0.01, \ \ f_{w}=S_{w},\ \lambda_{w}K=f_{w}D_{p},
\end{equation}
\begin{eqnarray}
F_{p}=0,\ F_{s}=2\pi^{2}D_{s}\sin{[\pi(x+y-2t)]}+2\pi(1-\phi)\cos{[\pi(x+y-2t)]}.
\end{eqnarray}
\end{subequations}
Under these parameters and some proper initial and boundary conditions considered in the domain $\Omega=[0, 1]\times[0, 1]$, one can also obtain exact solutions of pressure, velocity and saturation,
\begin{subequations}\label{eq4-5}
\begin{equation}
P(\mathbf{x}, t)=-\frac{\cos{[\pi(x+y-2t)]}}{5\pi}-\frac{(x+y)}{2},
\end{equation}
\begin{equation}
\mathbf{u}=(u_{x},\ u_{y})^{\intercal}=-D_{p} \nabla P=(1,\ 1)^{\intercal},
\end{equation}
\begin{equation}
S_{w}(\mathbf{x}, t)=\sin{[\pi(x+y-2t)]}.
\end{equation}
\end{subequations}

We first performed some simulations for the case of $\phi=1.0$ which has also been considered in the previous works \cite{Ohlberger1997,Mozolevski2013}, and presented the results at $T=0.2$ in Figs. 6-9 where the lattice size is $128\times 128$ and $c=10$. From these figures, one can observe that the numerical results are very close to their exact solutions, and the global relative errors of pressure, velocity and saturation are still less than $1.91\times10^{-3}$.

\begin{figure}
\includegraphics[width=2.5in]{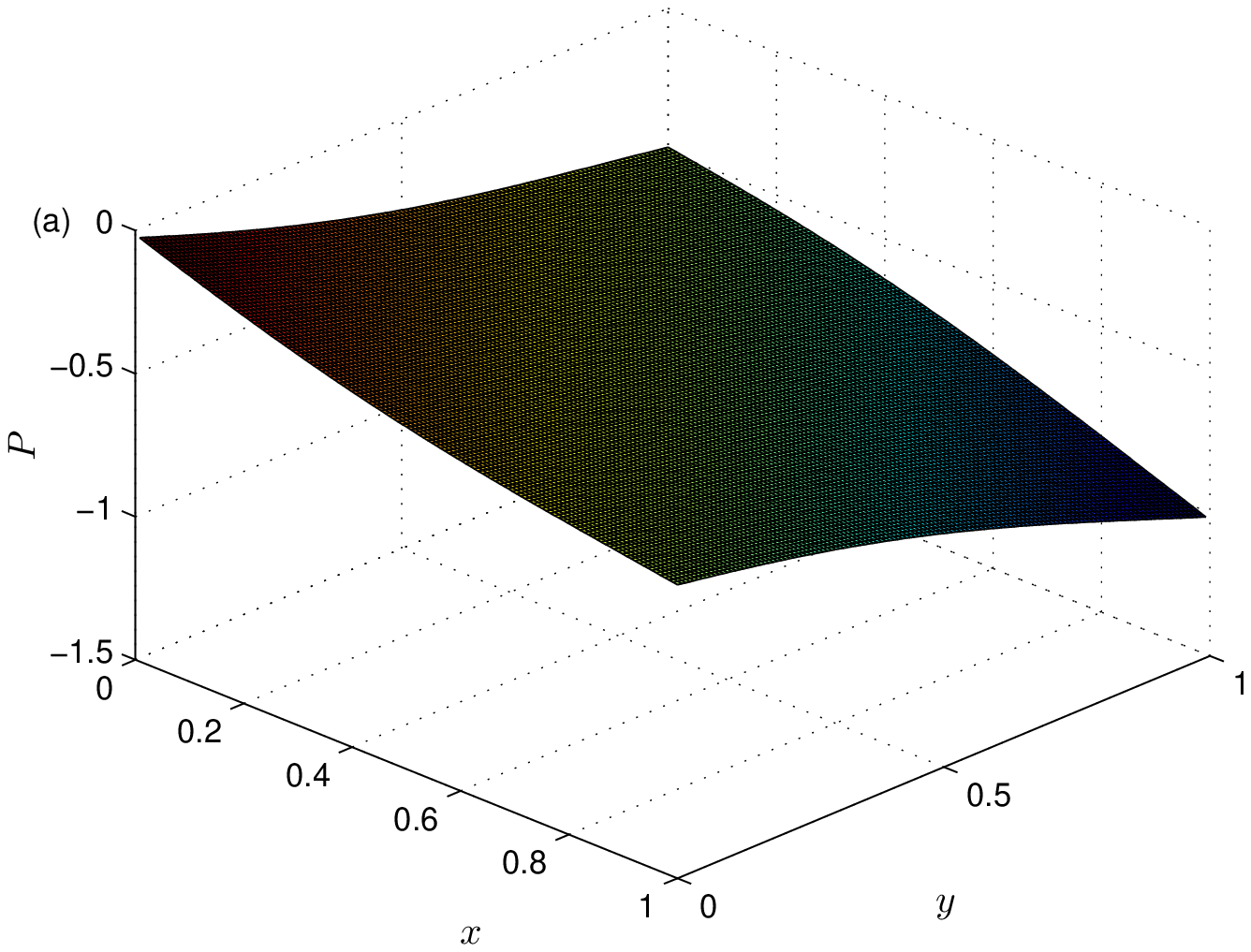}
\includegraphics[width=2.5in]{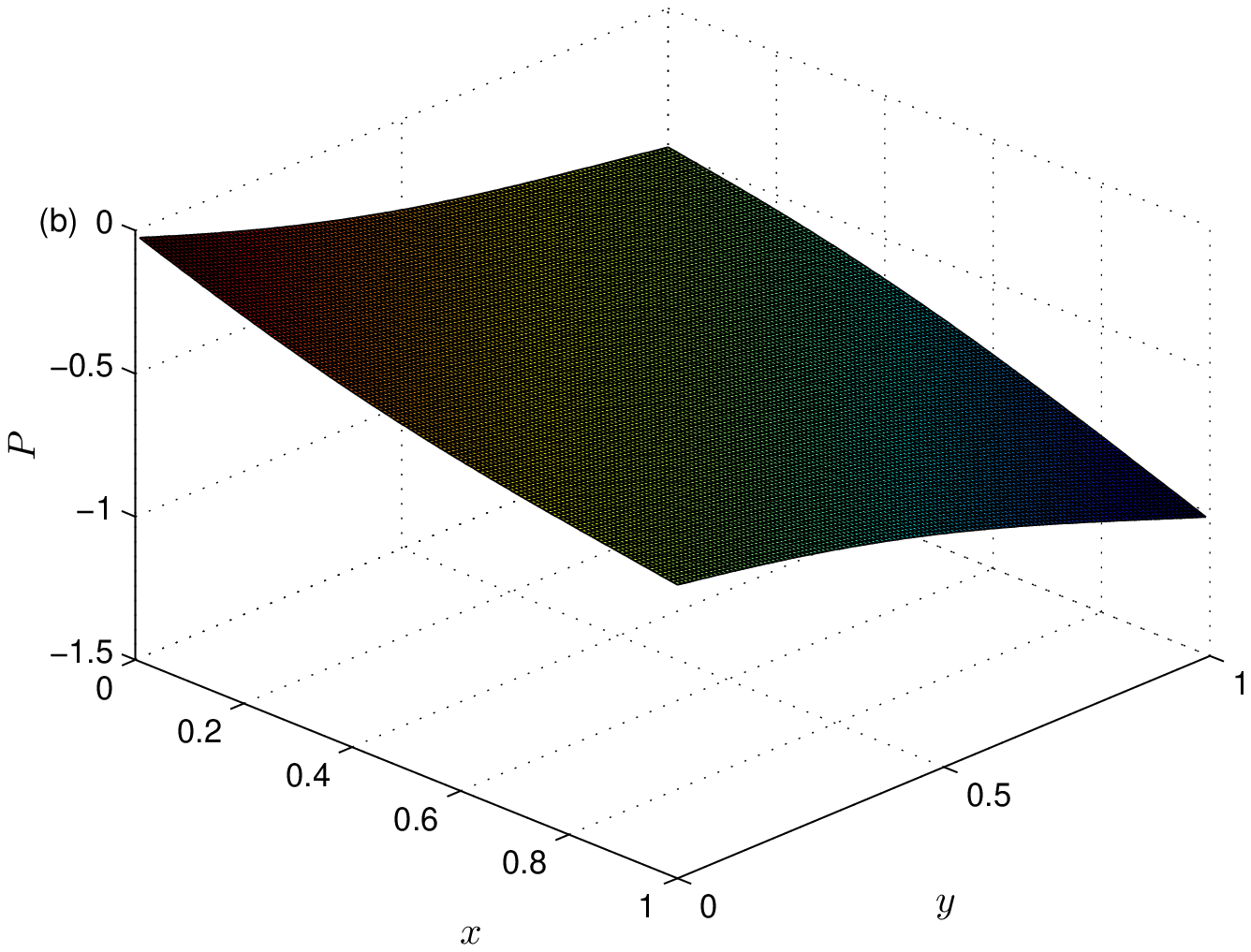}
\centering\caption{\label{fig:6} Distributions of pressure $P$ [(a): analytical solution, (b): numerical solution].}
\end{figure}

\begin{figure}
\includegraphics[width=2.5in]{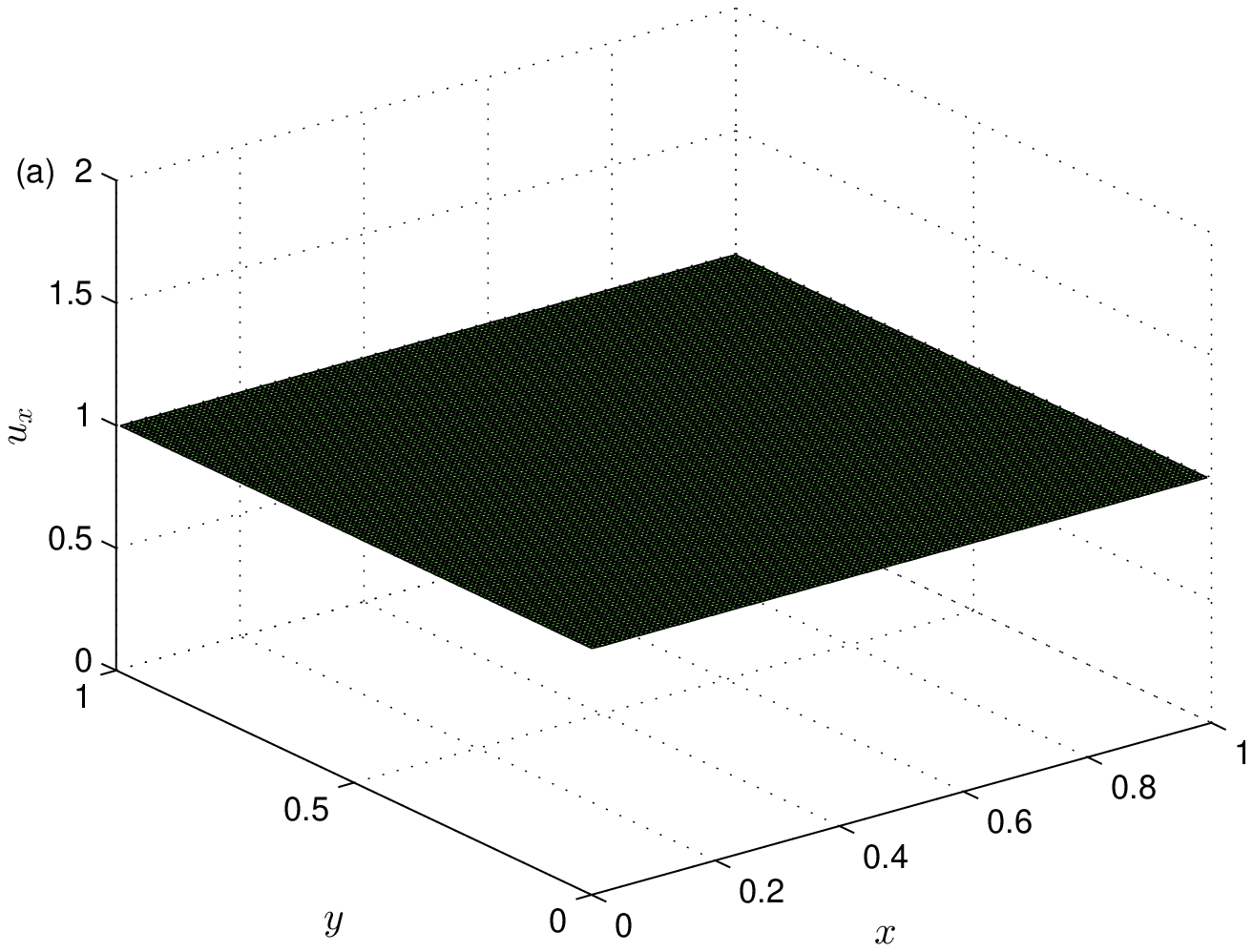}
\includegraphics[width=2.5in]{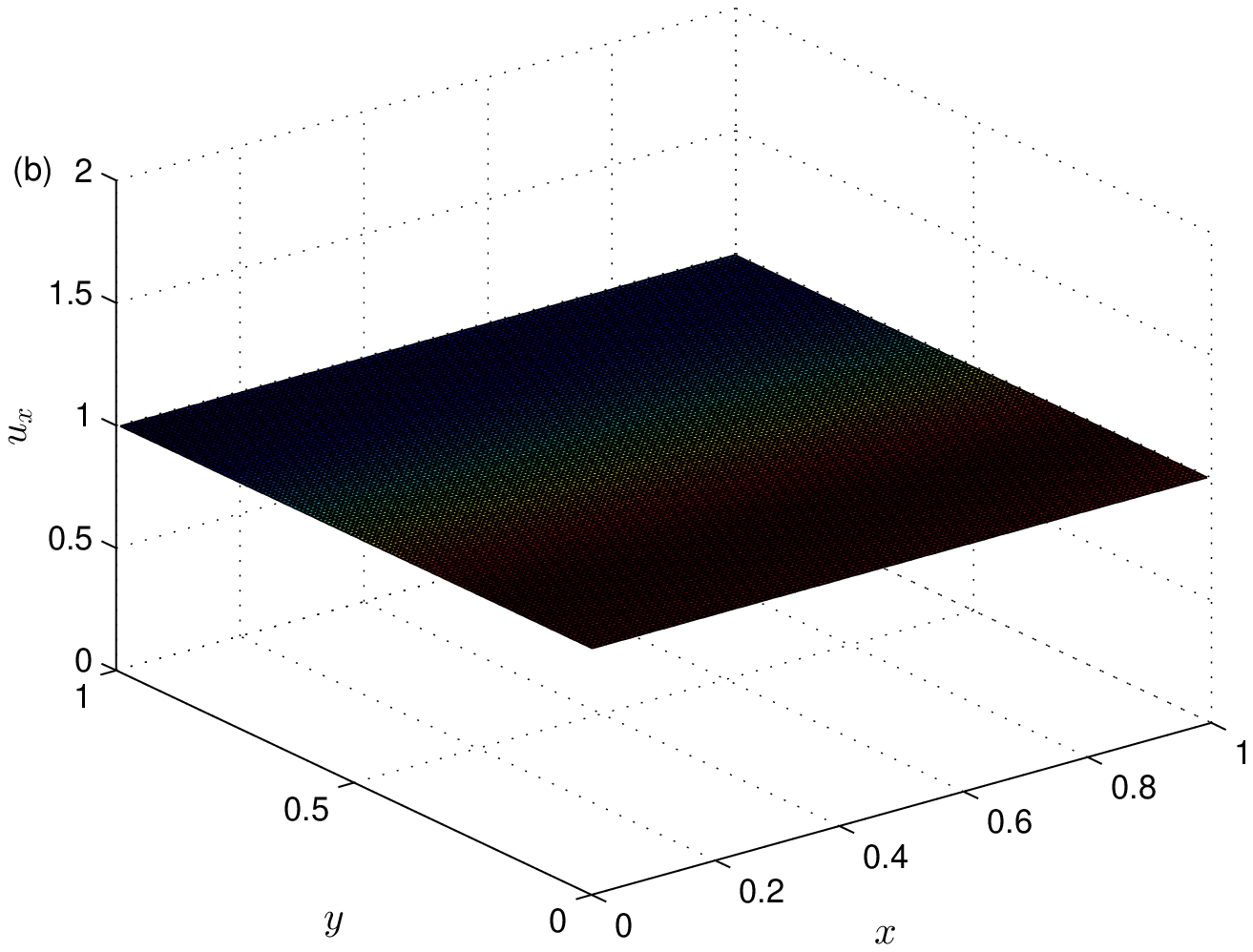}
\centering\caption{\label{fig:7} Distributions of velocity component $u_{x}$ [(a): analytical solution, (b): numerical solution].}
\end{figure}

\begin{figure}
\includegraphics[width=2.5in]{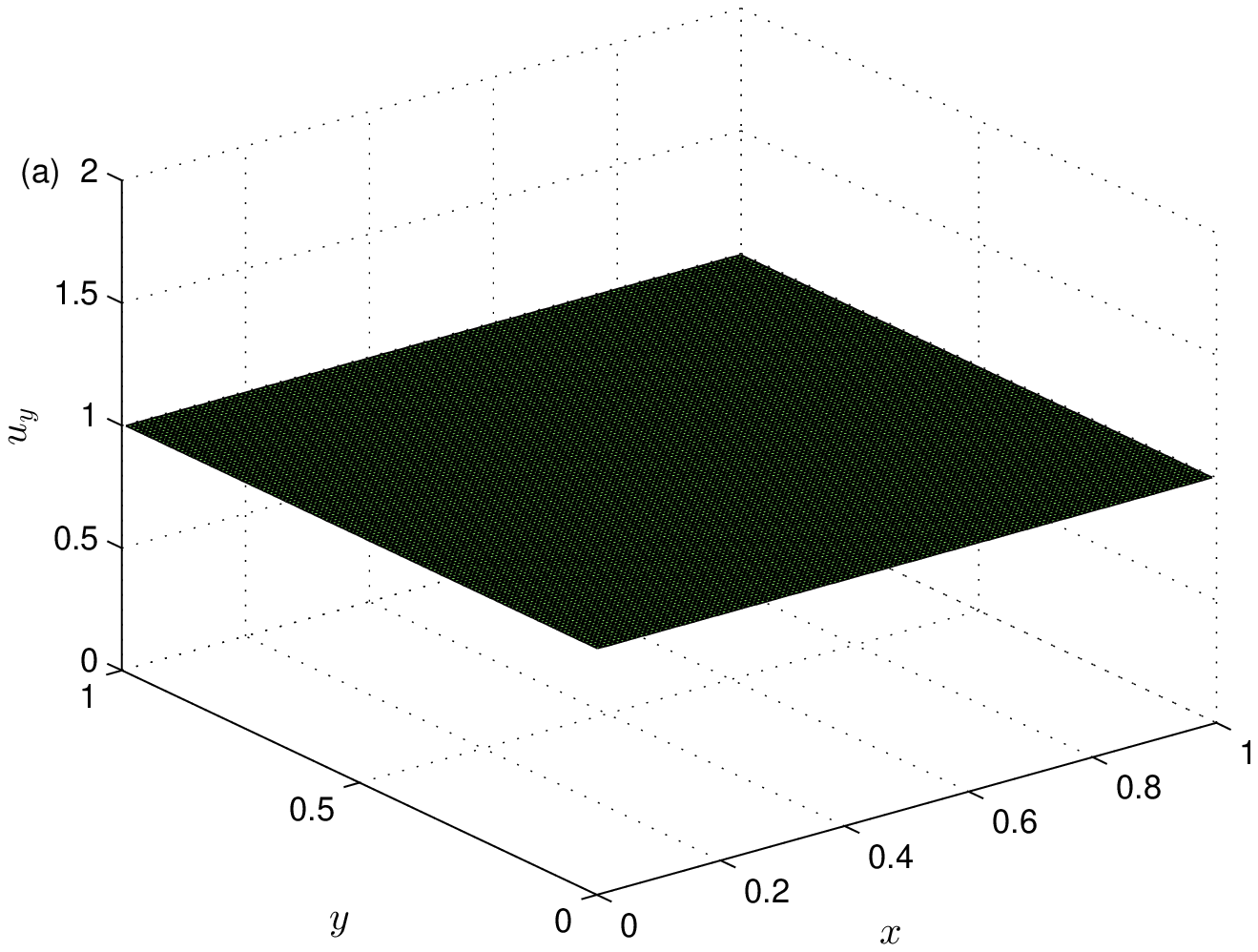}
\includegraphics[width=2.5in]{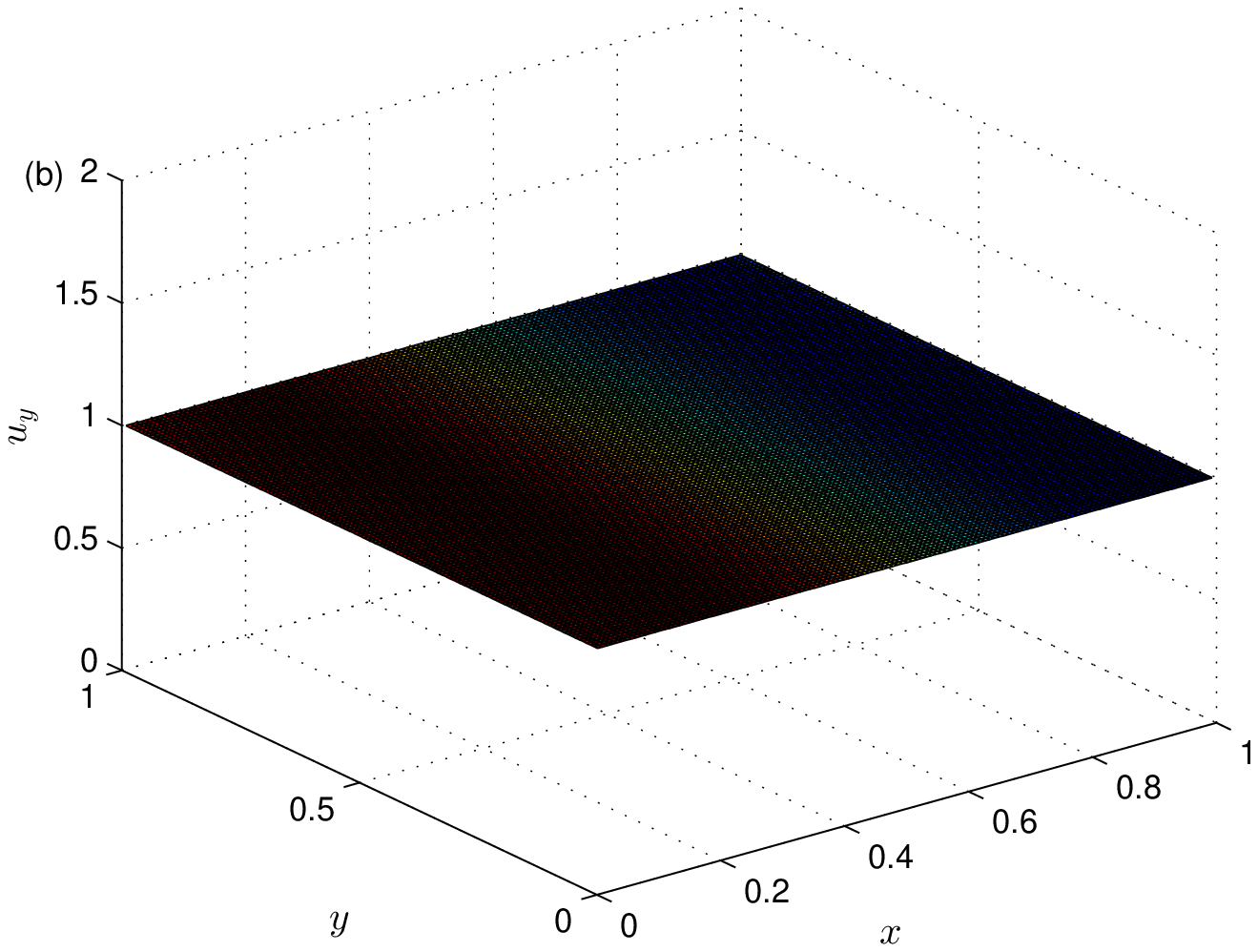}
\centering\caption{\label{fig:8} Distributions of velocity component $u_{y}$ [(a): analytical solution, (b): numerical solution].}
\end{figure}

\begin{figure}
\includegraphics[width=2.5in]{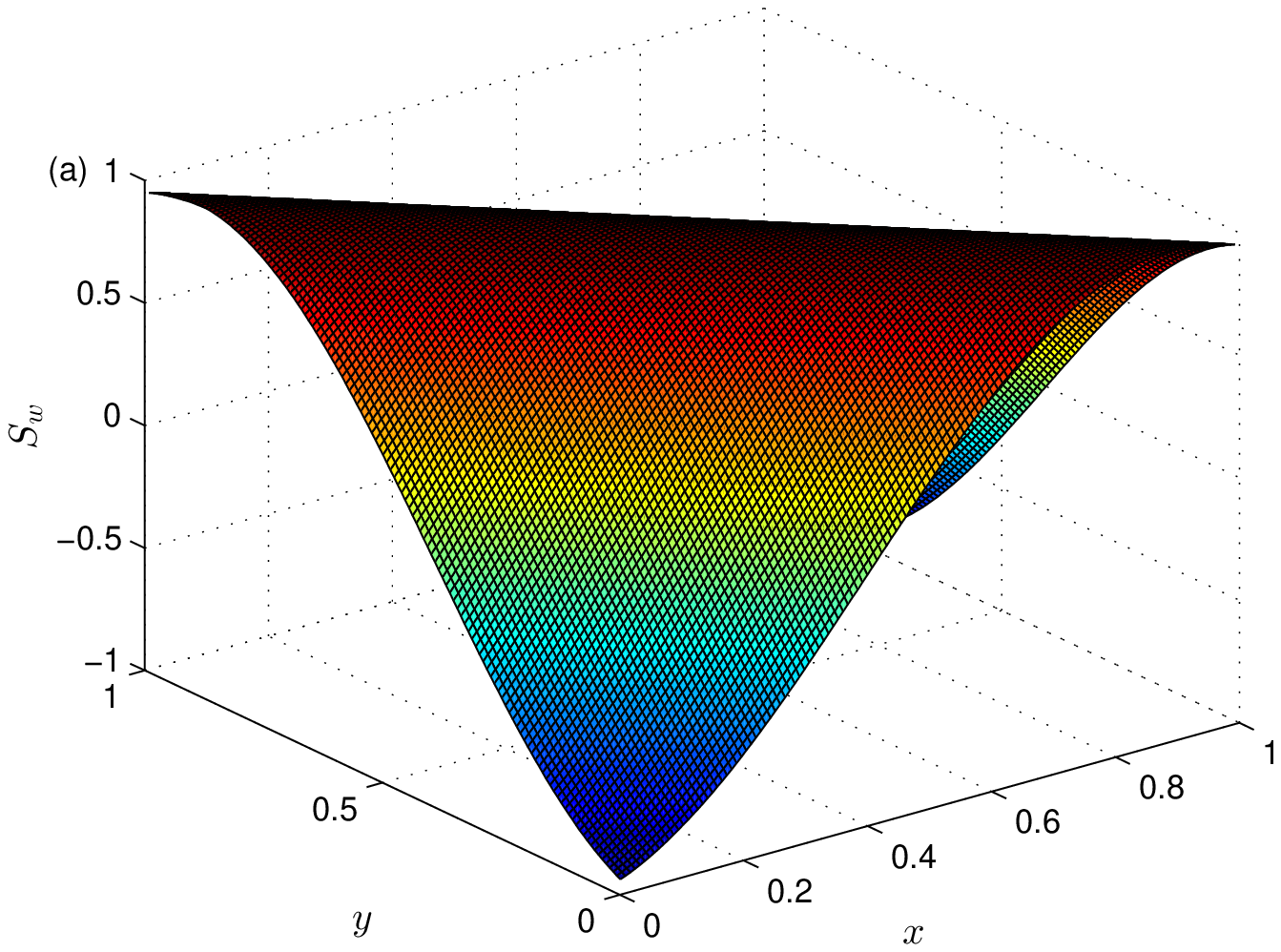}
\includegraphics[width=2.5in]{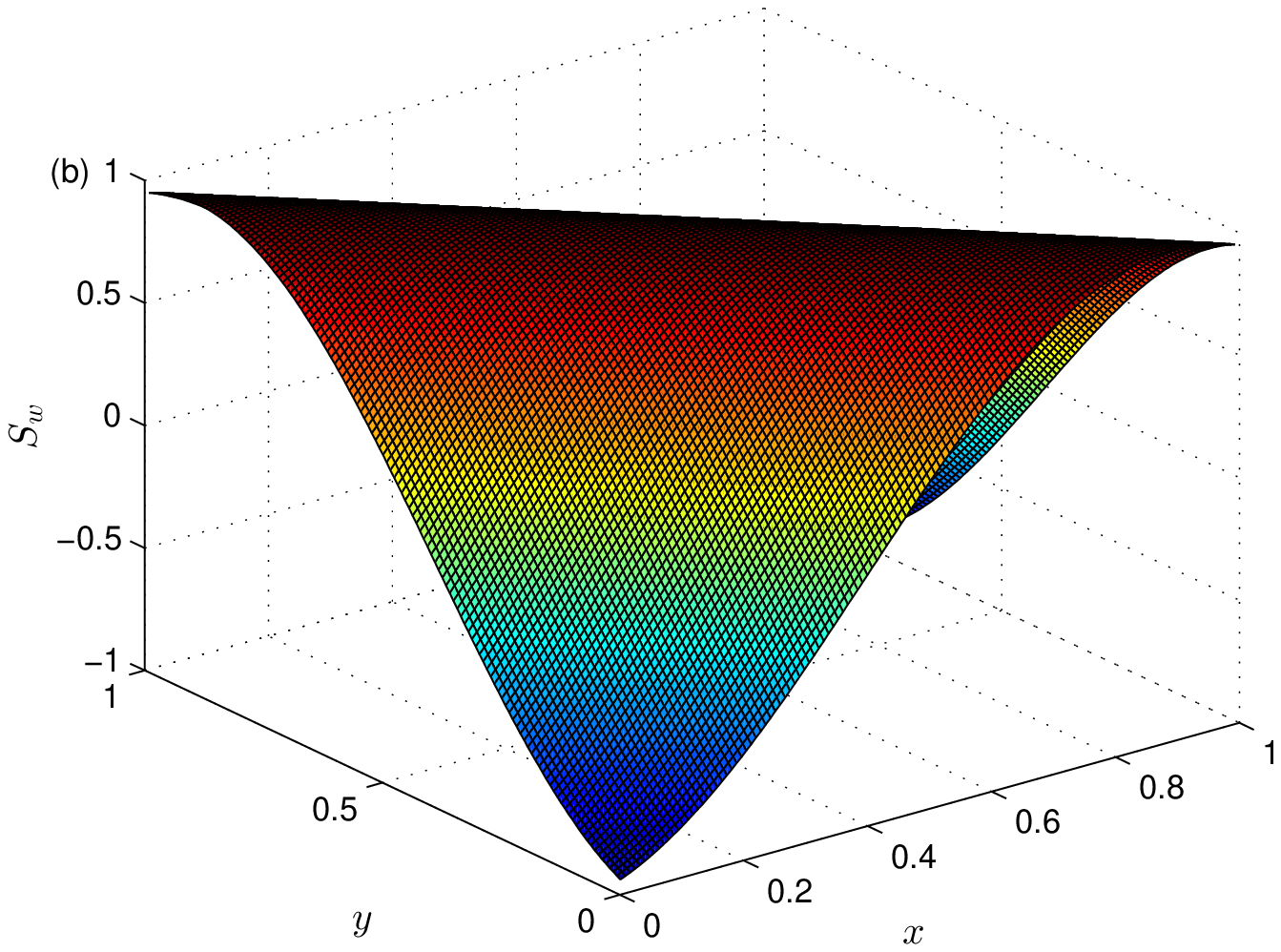}
\centering\caption{\label{fig:9} Distributions of saturation $S_{w}$ [(a): analytical solution, (b): numerical solution].}
\end{figure}

We also tested the convergence rate of present LB model with this example. To this end, we conducted some simulations with the same physical parameters mentioned above and different lattice sizes, and presented the global relative errors of pressure, velocity and saturation in Fig. 10. As shown in this figure, the present LB model also has a second-order convergence rate even for this coupled problem.

\begin{figure}
\includegraphics[width=3in]{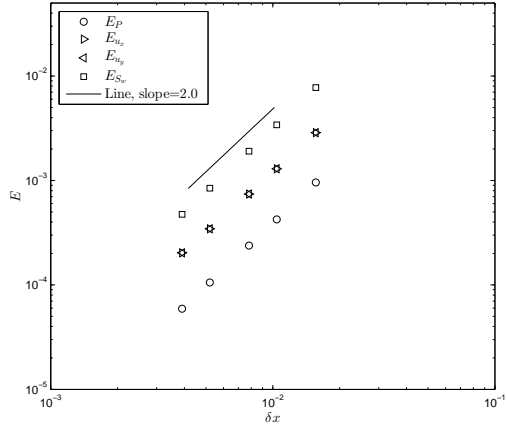}
\centering\caption{\label{fig:10} The global relative errors of pressure, velocity and saturation at different lattice sizes ($\delta x=L/256, \ L/192,\ L/128,\ L/96,\ L/64$, $L=1.0$). The slope of inserted line is 2.0, which indicates that the present LB model has a second-order convergence rate.}
\end{figure}

It should be noted that all above simulations are conducted only for the case of $\phi=1.0$, while for the two-phase flows in porous media, $\phi$ is less than 1. For this reason, we also carried out some simulations for the case of $\phi=0.5$. To ensure that our simulations are stable, here we also considered different values of $\beta$ ($\beta=0.25,\ 0.5, \ 0.75, \ 1.0$) under the condition of $\phi/2\leq\beta\leq2\phi$, and only the errors of pressure, velocity and saturation are presented in Fig. 11 since the distributions of pressure, velocity and saturation are similar to those in Figs. 6-9. As seen from this figure, the parameter $\beta$ indeed has an apparent influence on numerical results, the errors of pressure, velocity and saturation decrease with the increase of $\beta$, but it does not affect the second-order convergence rate in space.

\begin{figure}
\includegraphics[width=2.5in]{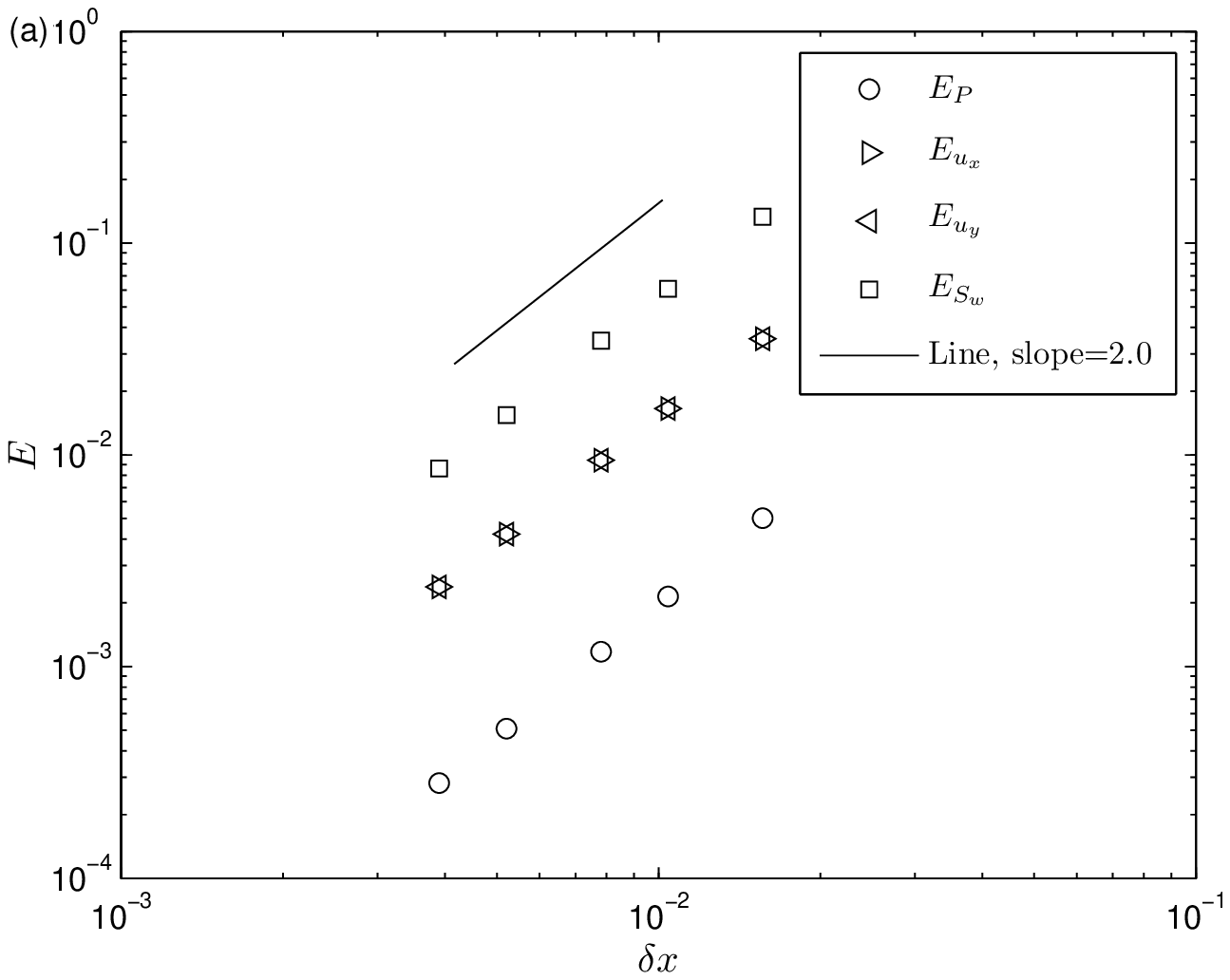}
\includegraphics[width=2.5in]{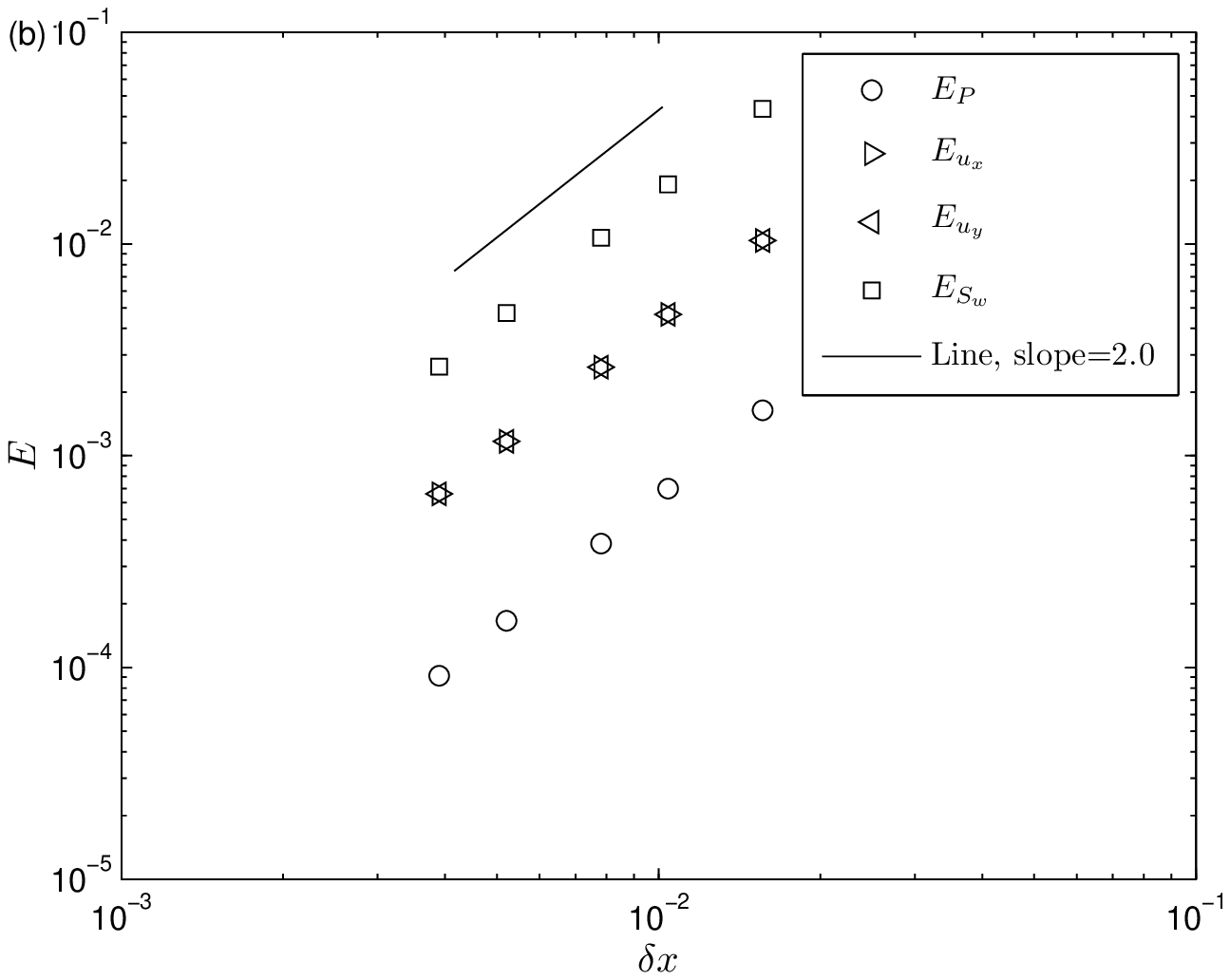}
\includegraphics[width=2.5in]{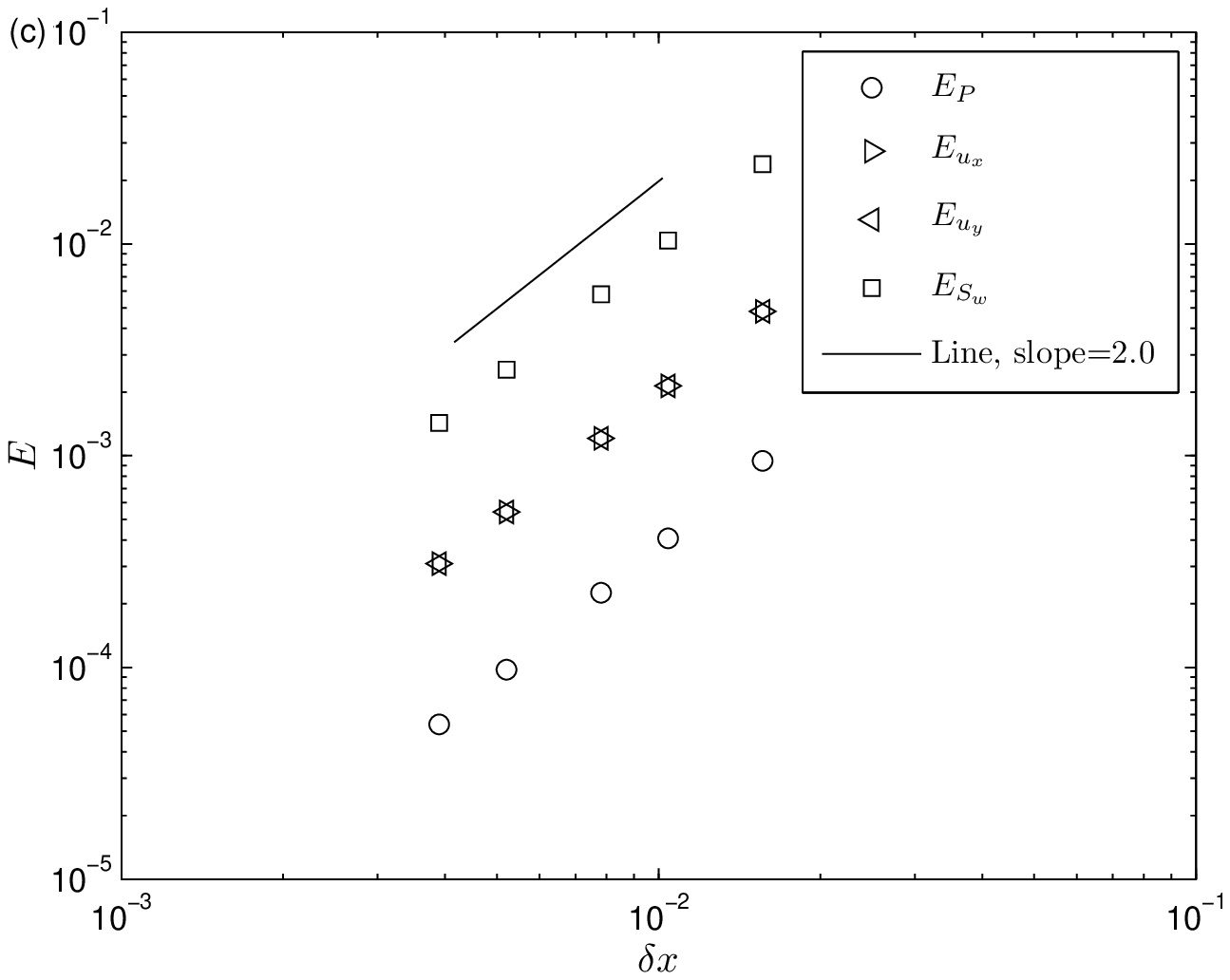}
\includegraphics[width=2.5in]{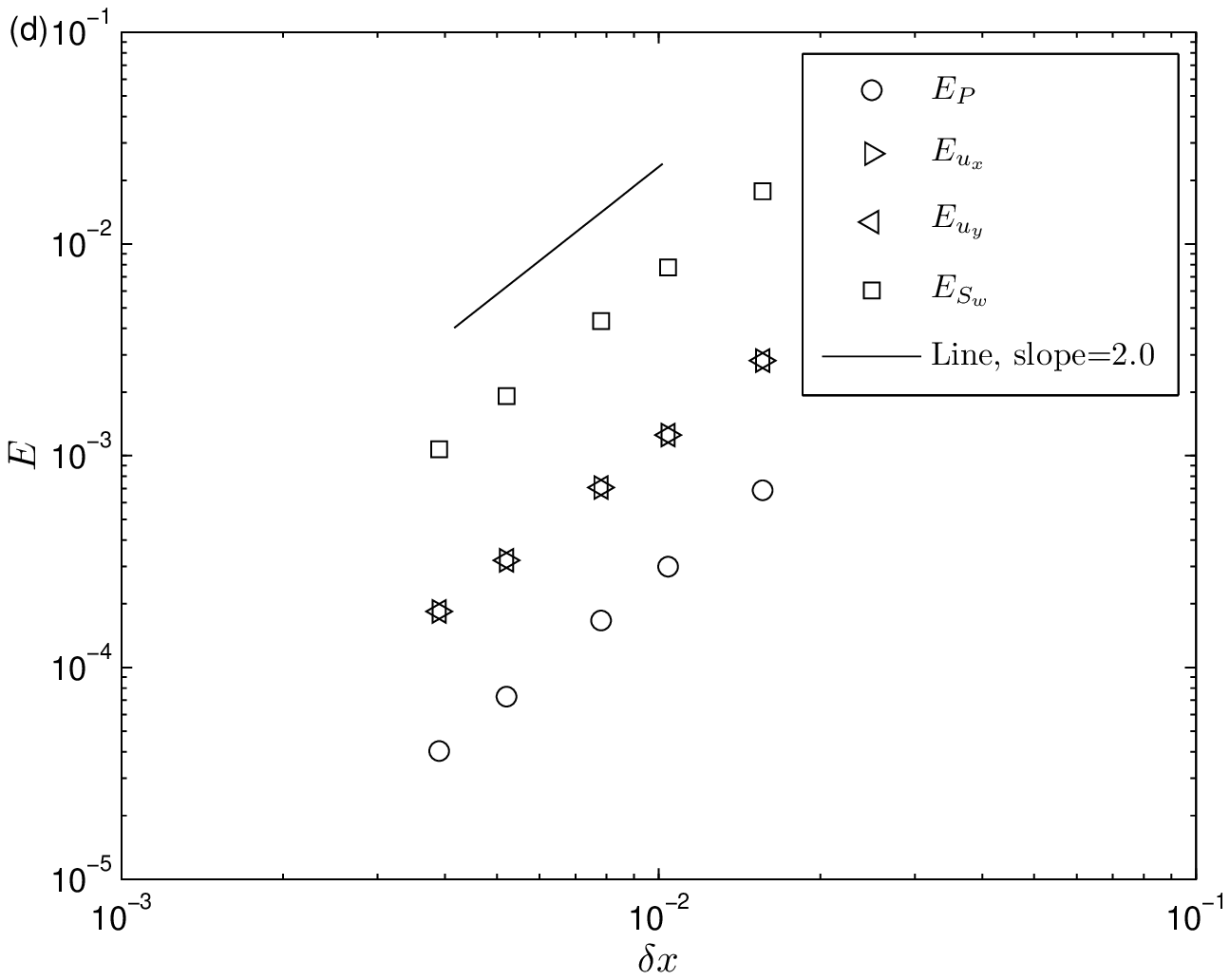}
\centering\caption{\label{fig:11} The global relative errors of pressure, velocity and saturation at different values of $\beta$ [(a): $\beta=0.25$, (b): $\beta=0.5$, (c): $\beta=0.75$, (d): $\beta=1.0$] and different lattice sizes ($\delta x=L/256, \ L/192,\ L/128,\ L/96,\ L/64$, $L=1.0$). The slope of inserted line is 2.0, indicating that the present LB model has a second-order convergence rate..}
\end{figure}

\subsection{Example 3: The classical five-Spot problem}

\begin{figure}
\includegraphics[width=4in]{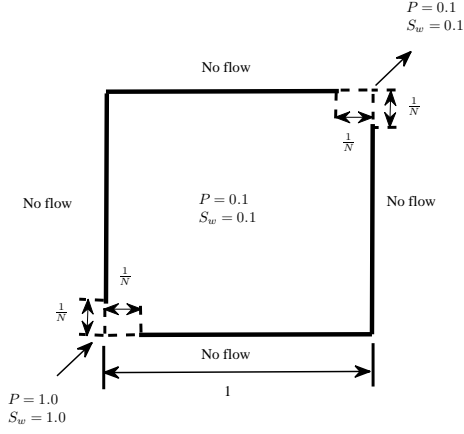}
\centering\caption{\label{fig:12} Schematic of the five-spot problem.}
\end{figure}

The last problem we considered is the classical five spot problem \cite{Chen2006,Cao2011,Lewis1984,Kukreti1989}, which is more complicated, and there is no analytical solution available. Based on the previous works \cite{Cao2011,Kukreti1989}, the problem (see Fig. \ref{fig:12}) can be described by the following simplified mathematical model,
\begin{subequations}\label{eq4-6}
\begin{equation}
\nabla\cdot(D_{p}\nabla P)=0, \ \ \mathbf{u}=-D_{p}\nabla P,
\end{equation}
\begin{equation}
\phi\frac{\partial S_{w}}{\partial t} = \nabla\cdot(D_{s}\nabla S_{w}) + \nabla\cdot(\lambda_{w}K\nabla P),
\end{equation}
\end{subequations}
where the physical parameters appeared in Eq.~(\ref{eq4-6}) are listed in Table 1.

\begin{table*}
\caption{Some physical parameters used in example 3.}
\begin{center}
\begin{tabular}{cc}
\hline
  Parameter  & Value \\
\hline
 $\phi$  & 1.0 \\
 $\mu_{w}$, $\mu_{n}$ & $1.0\times10^{-3}$ \\
 $K$ & $1.0\times10^{-5}$ \\
 $D_{s}$ & $1.0\times10^{-2}$ \\
\hline
\end{tabular}
\end{center}
\end{table*}
To determine the fractional flow function $\lambda_{w}$ and total mobility $\lambda_{t}$, the following quadratic relative permeabilities are adopted,
\begin{equation}
k_{rw}=S_{w}^{2}, \ \ \ k_{rn}=(1.0-S_{w})^{2}.
\end{equation}

\begin{figure}
\includegraphics[width=2.5in]{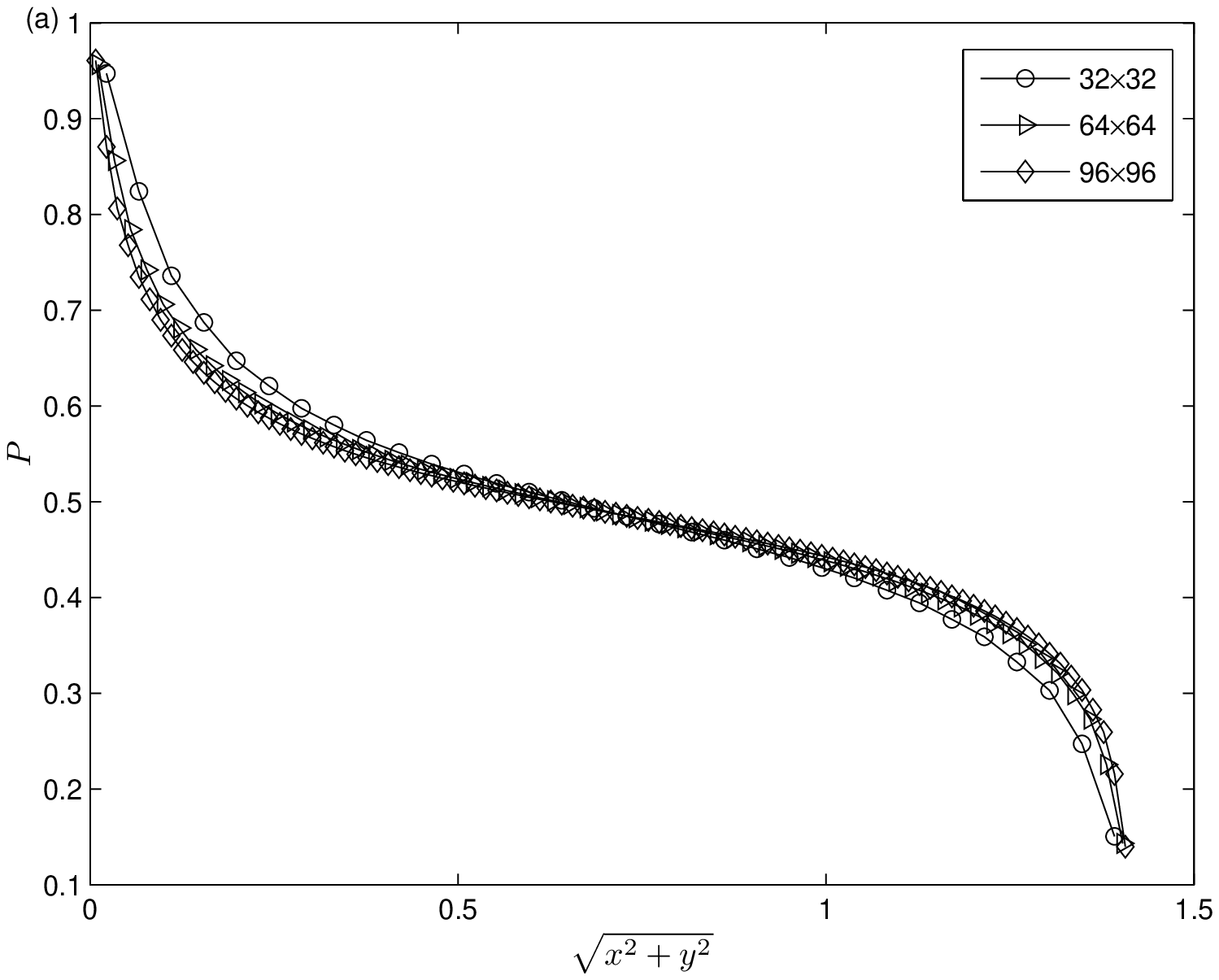}
\includegraphics[width=2.5in]{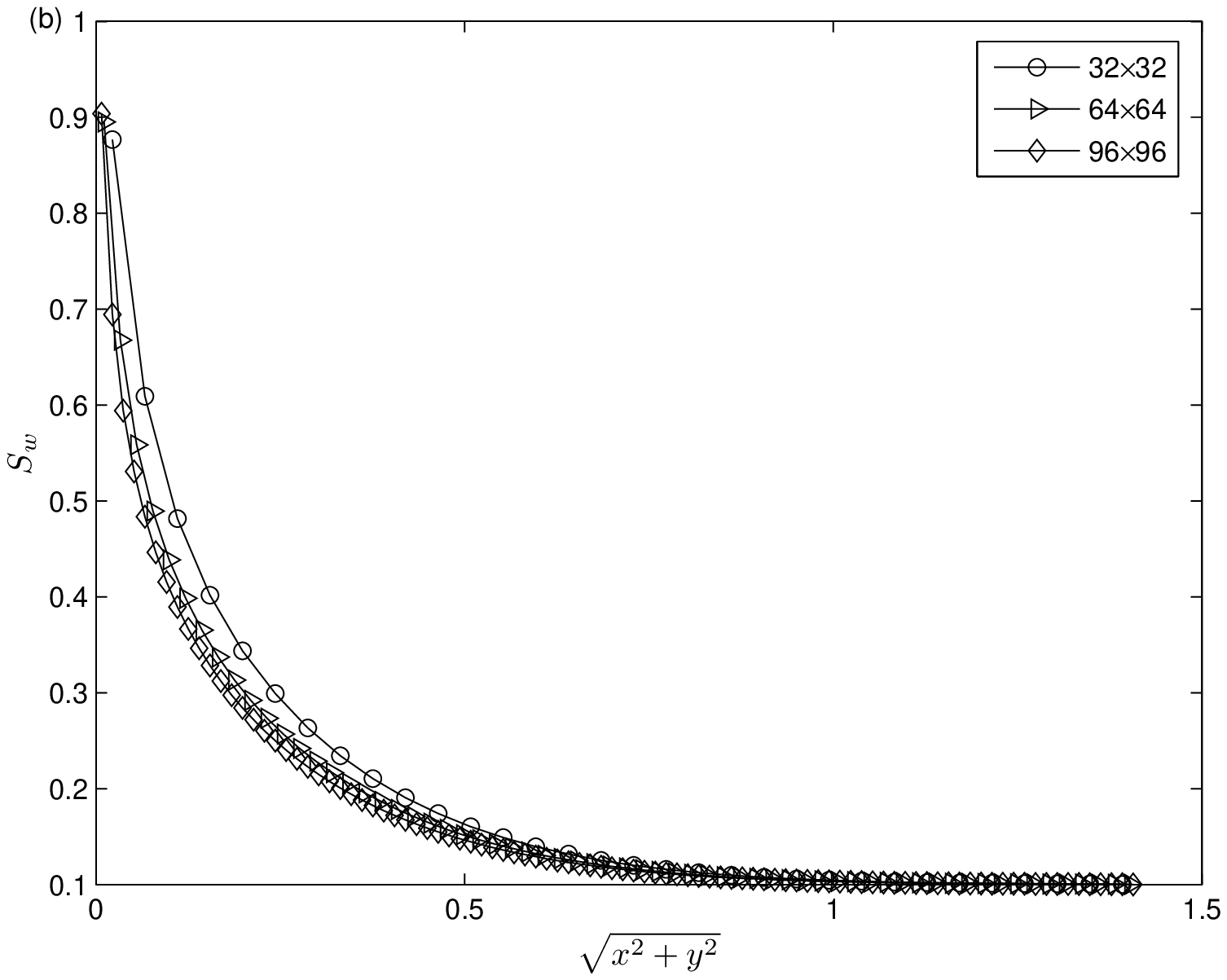}
\centering\caption{\label{fig:13} Distributions of saturation $s_{w}$ and pressure $P$ alone the diagonal line.}
\end{figure}

The physical domain of the problem is $\Omega=[0,\ 1]\times[0,\ 1]$, and the initial and boundary conditions are depicted in Fig. \ref{fig:12}. We note that there is no exact solution to this problem, and for this reason, a grid-independence test is first conducted. We carried out some simulations under three different lattice sizes $N\times N=32\times 32$, $64\times 64$ and $96\times 96$, and presented distributions of saturation and pressure alone the diagonal line in Fig. \ref{fig:13} where $T=10$. As shown in this figure, the lattice size $96\times 96$ is fine enough, and can give grid-independence results.

We then performed some simulations with the lattice size $96\times 96$, and presented the distributions of saturation, pressure and streamline at different time in Figs. 14-16. From these figure, one can find that the present results are similar to those reported in some previous works \cite{Cao2011,Kukreti1989}. Besides, we also note that the streamlines at $T=10,\ 20, 30$ and 40 are very close to each other (see Fig. \ref{fig:16}), which is caused by the small changes of velocity or pressure gradient at these different time, as seen clearly from Fig. \ref{fig:14}.

\begin{figure}
\includegraphics[width=2.5in]{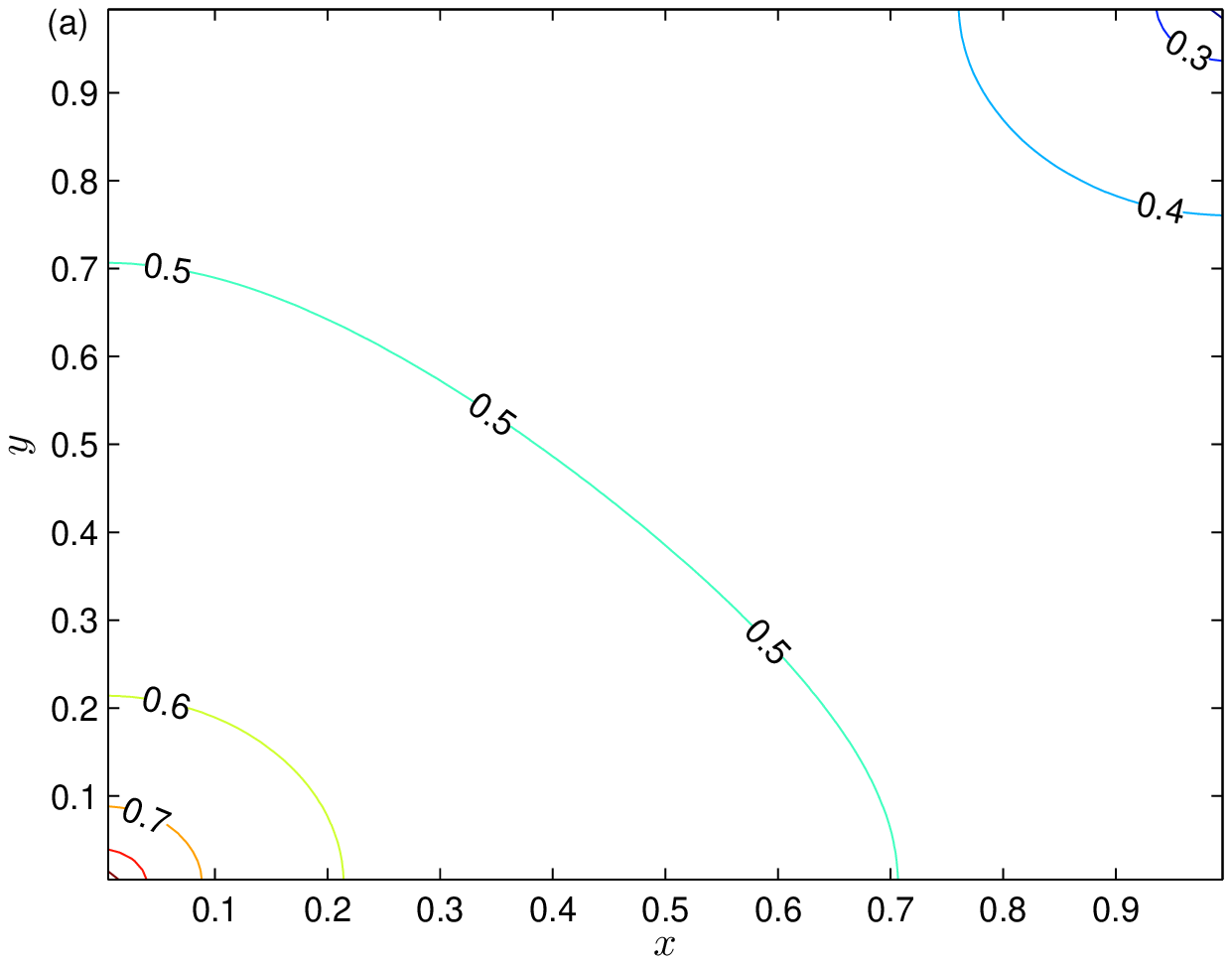}
\includegraphics[width=2.5in]{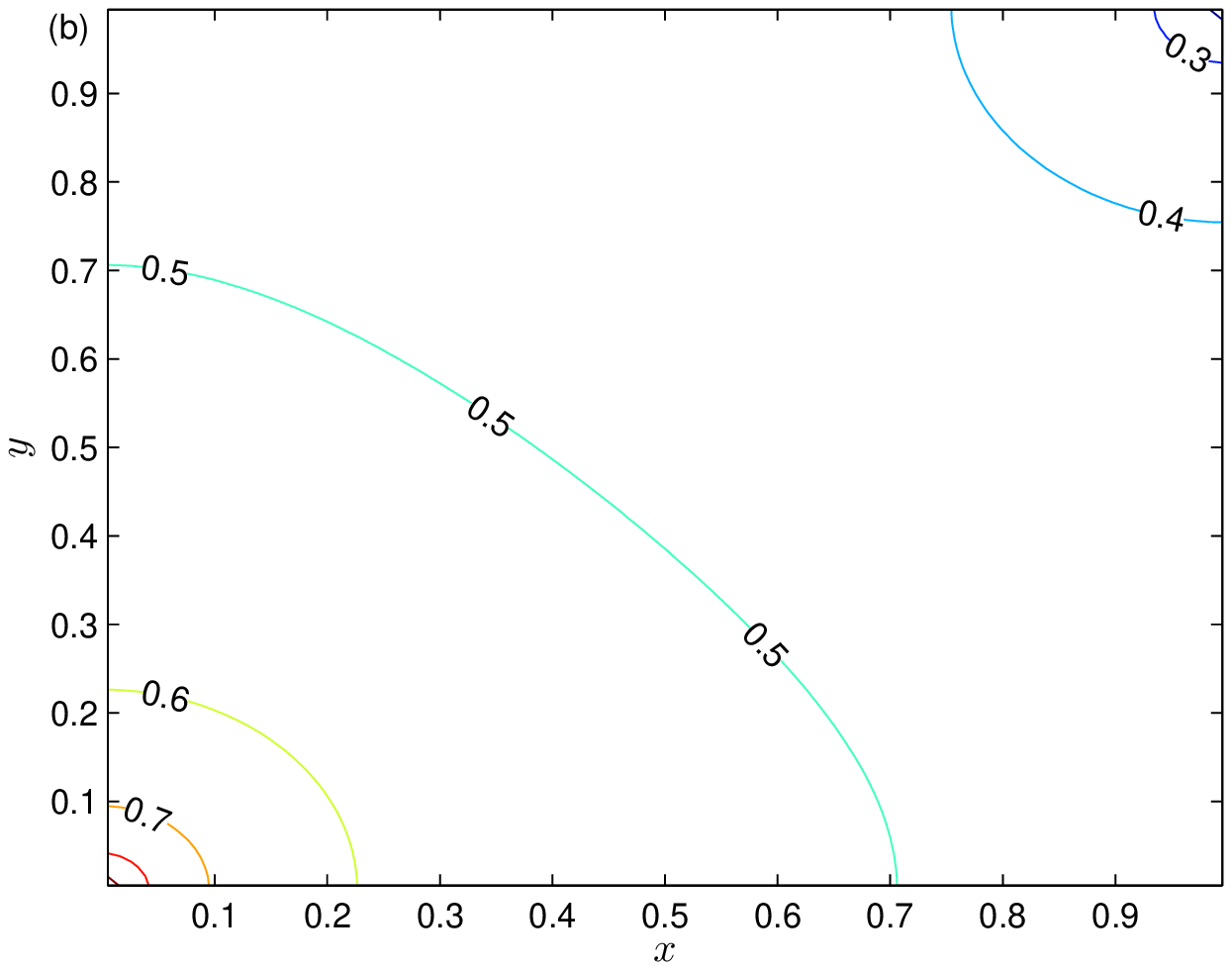}
\includegraphics[width=2.5in]{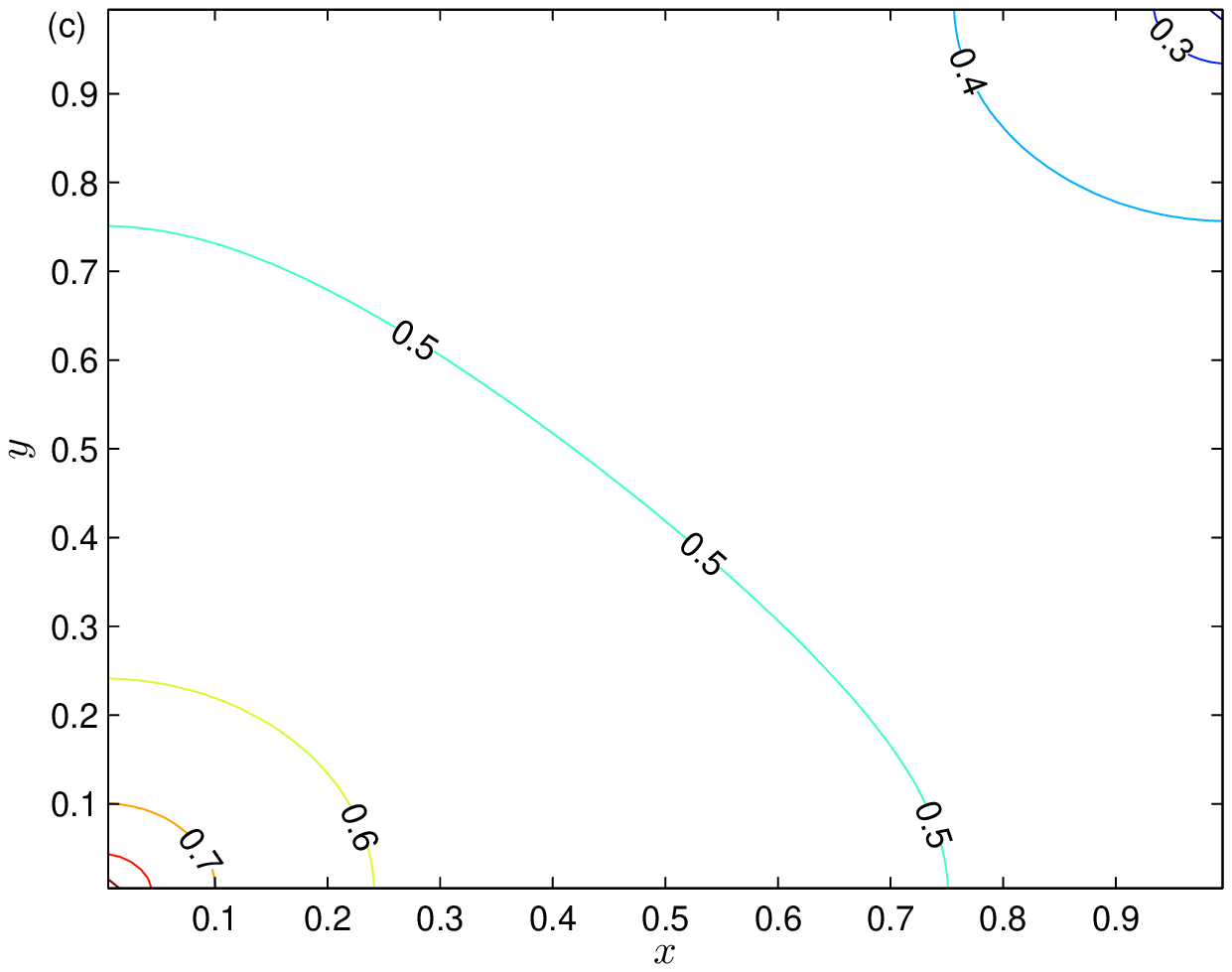}
\includegraphics[width=2.5in]{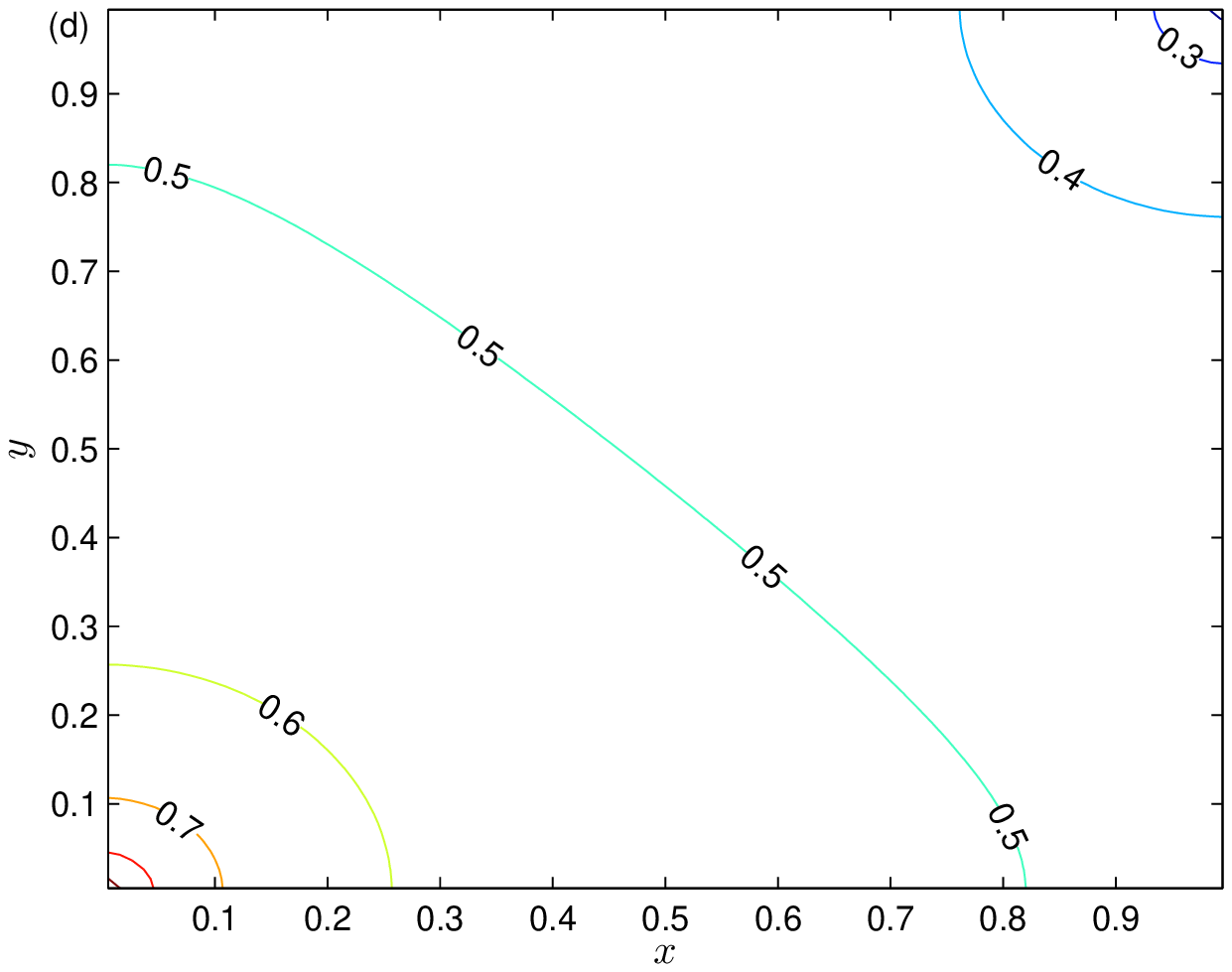}
\centering\caption{\label{fig:14} Contour lines of pressure $P$ [(a): $T=10$, (b): $T=20$, (c): $T=30$, (d): $T=40$].}
\end{figure}

\begin{figure}
\includegraphics[width=2.5in]{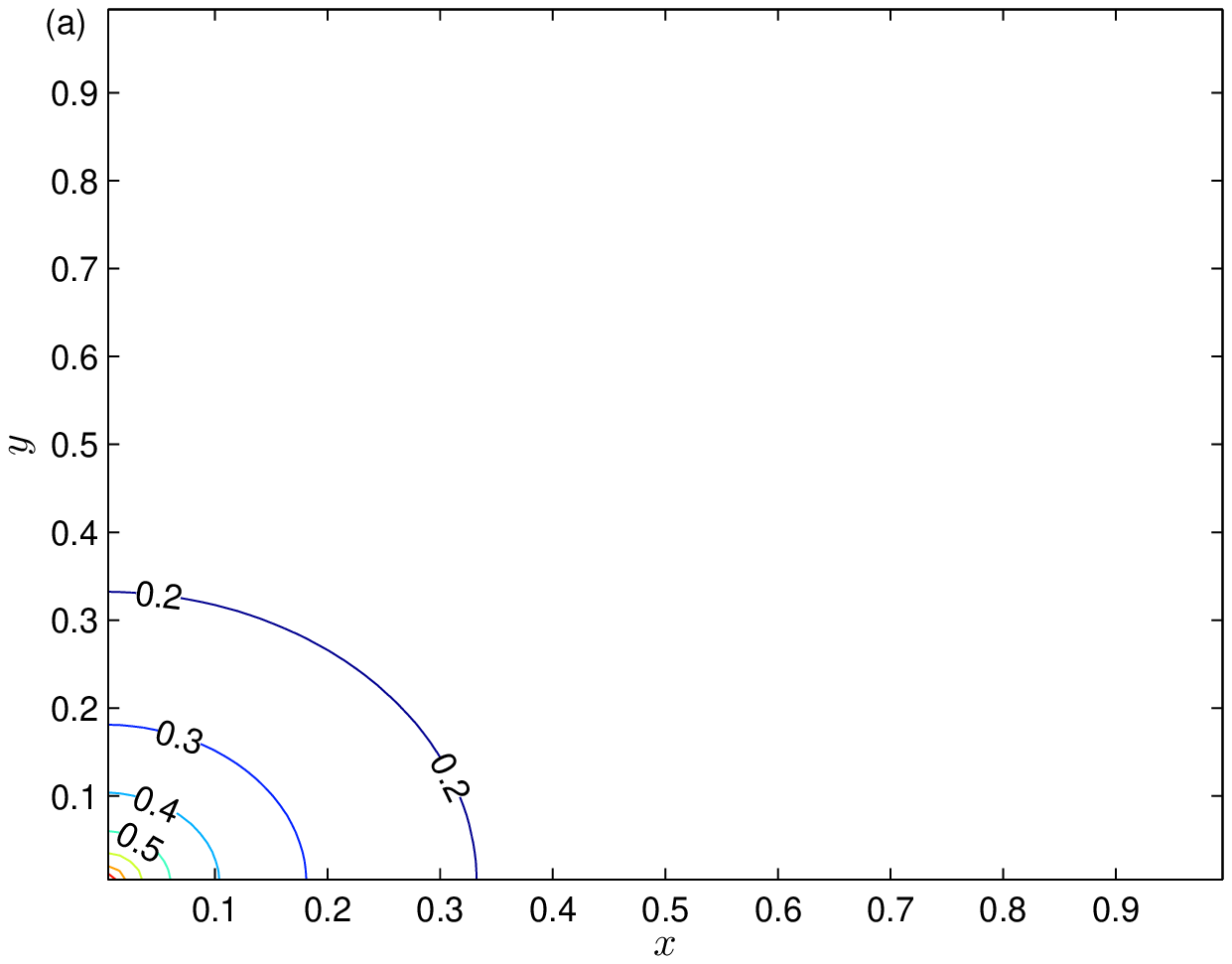}
\includegraphics[width=2.5in]{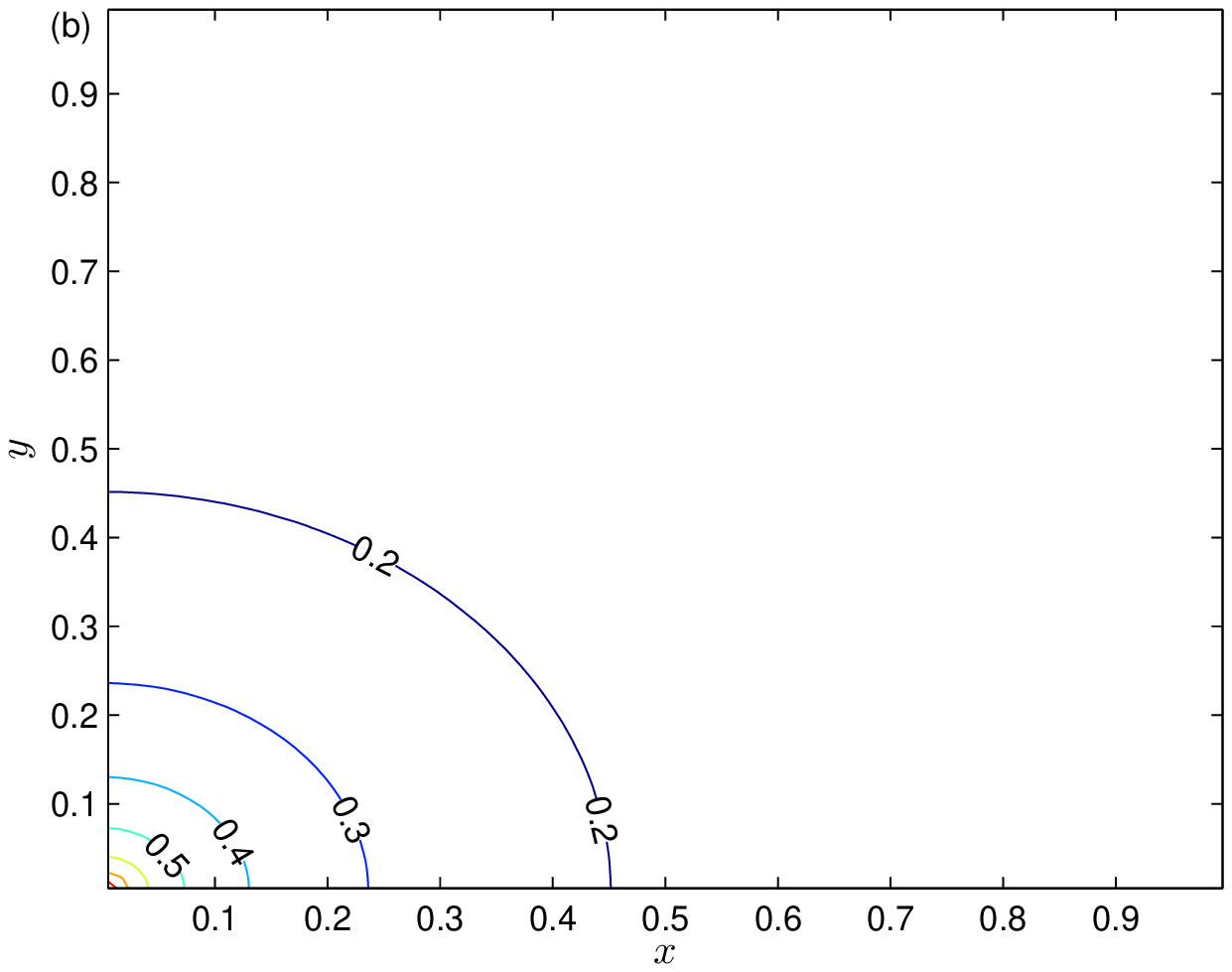}
\includegraphics[width=2.5in]{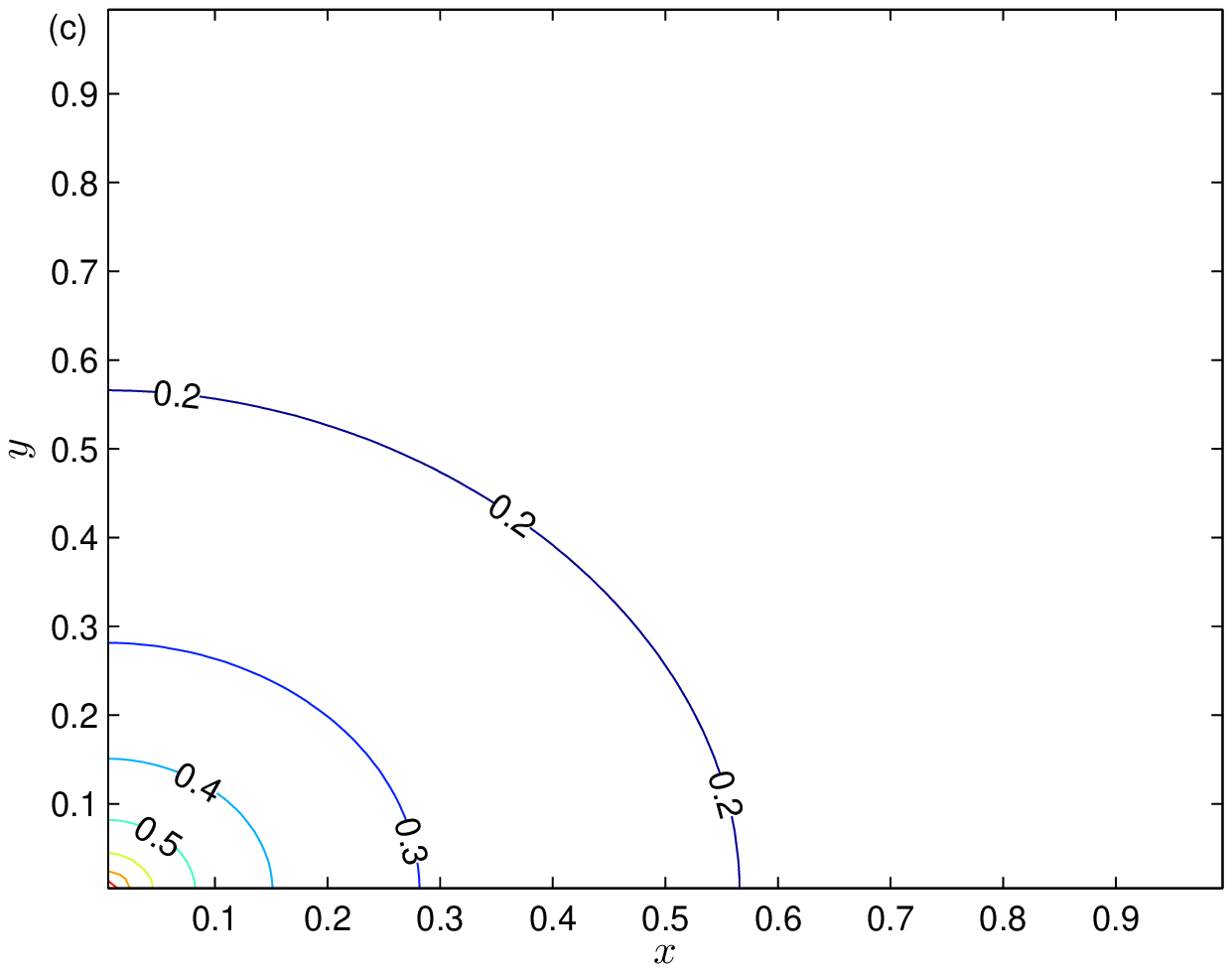}
\includegraphics[width=2.5in]{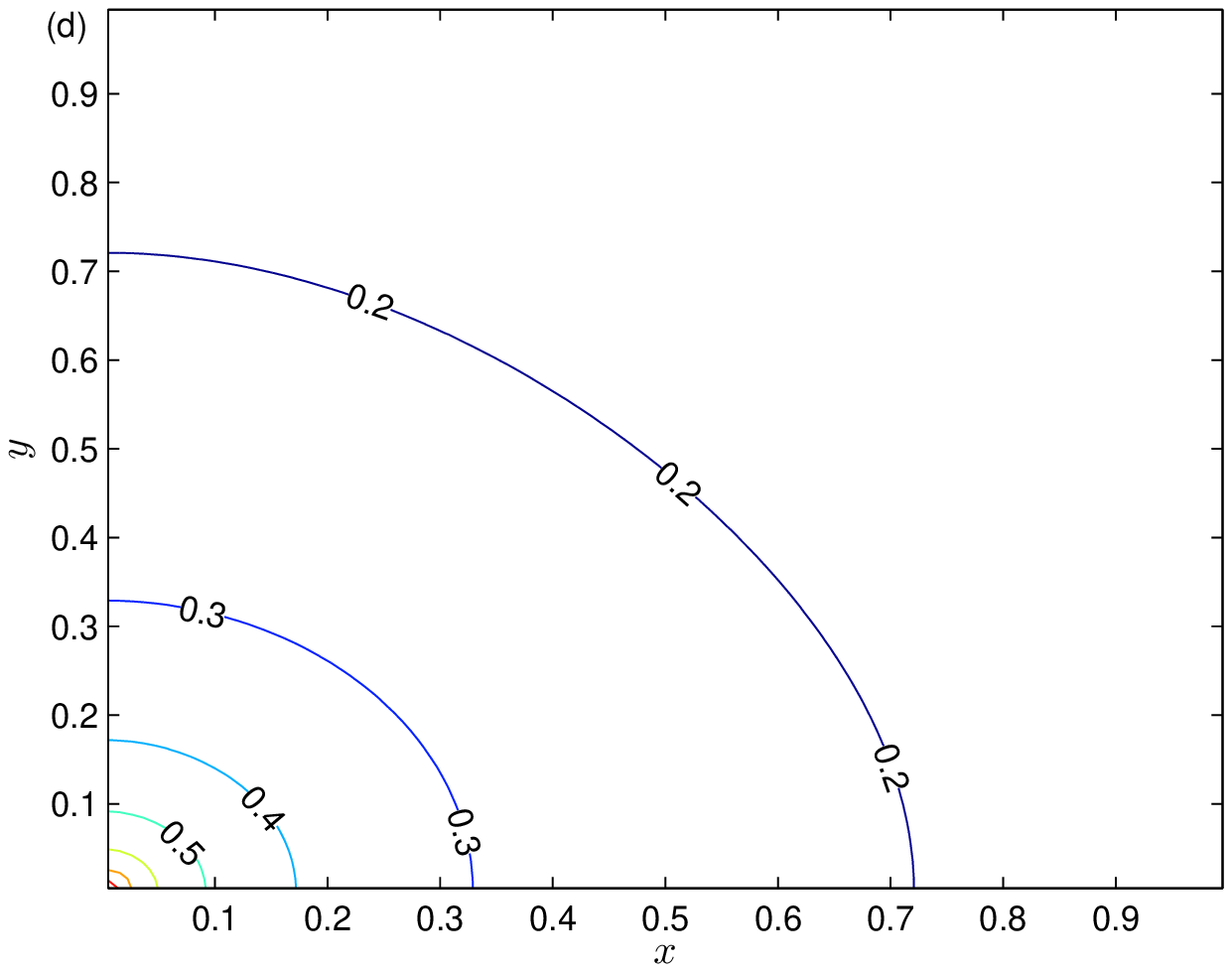}
\centering\caption{\label{fig:14} Contour lines of saturation $S_{w}$ [(a): $T=10$, (b): $T=20$, (c): $T=30$, (d): $T=40$].}
\end{figure}

\begin{figure}
\includegraphics[width=2.5in]{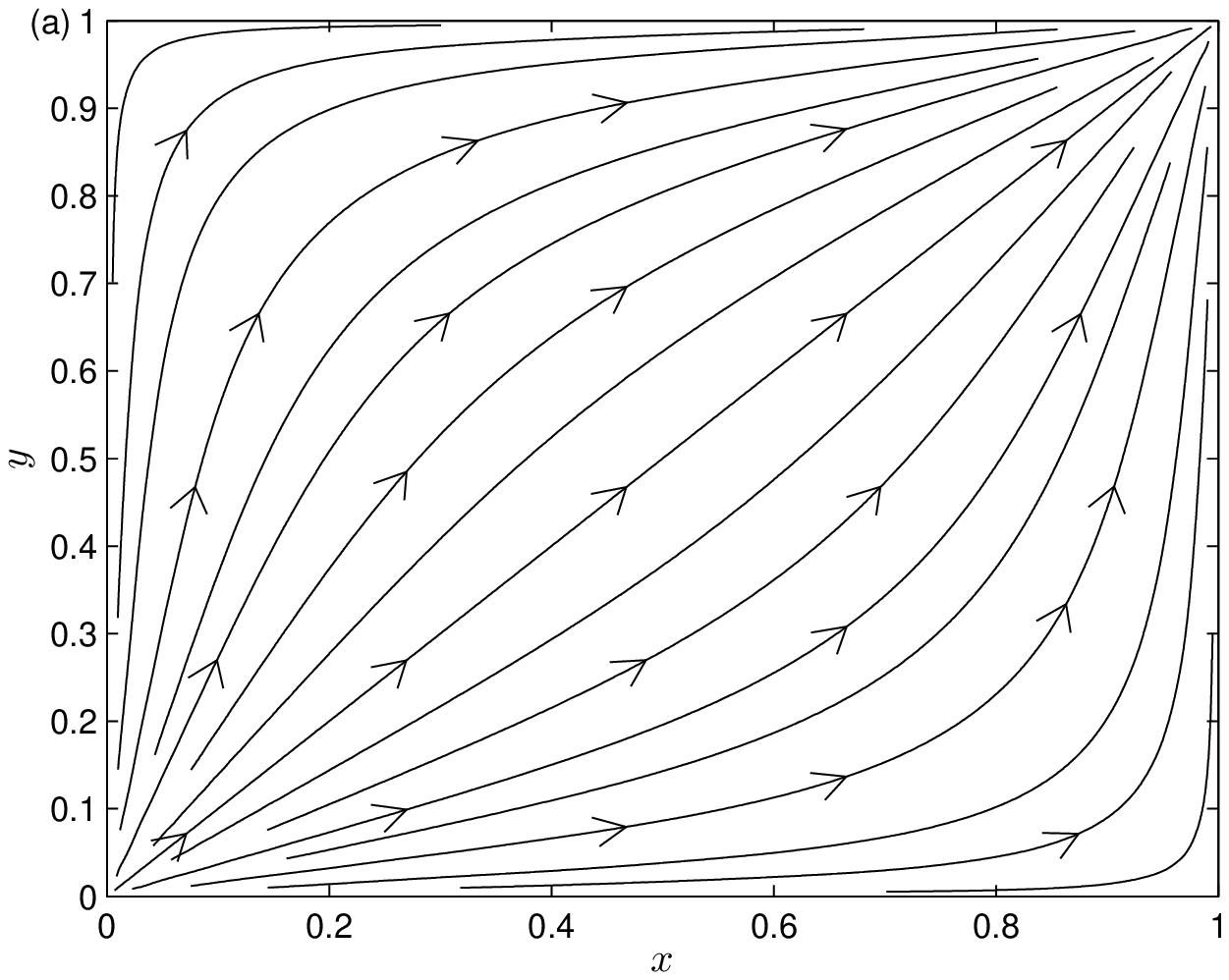}
\includegraphics[width=2.5in]{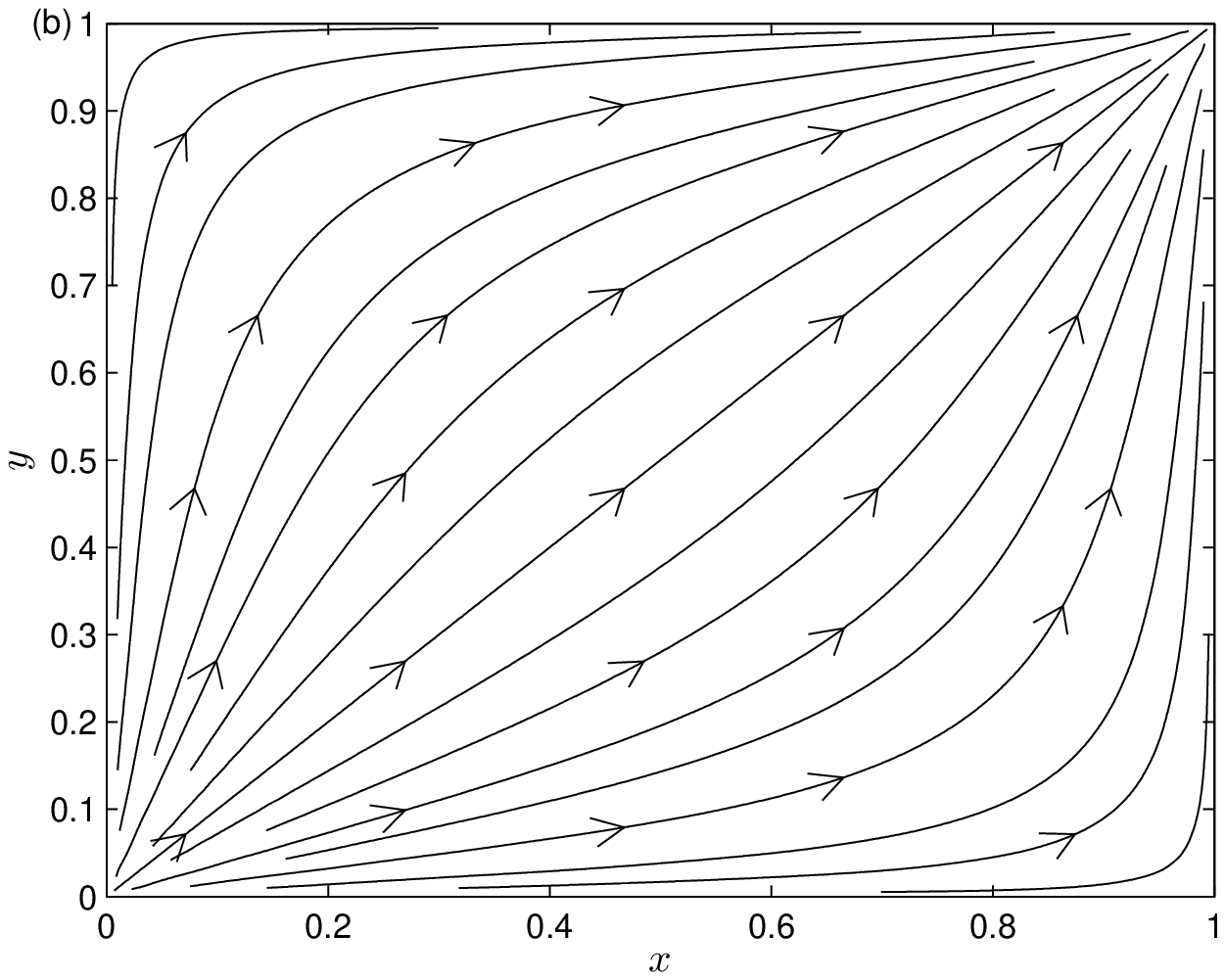}
\includegraphics[width=2.5in]{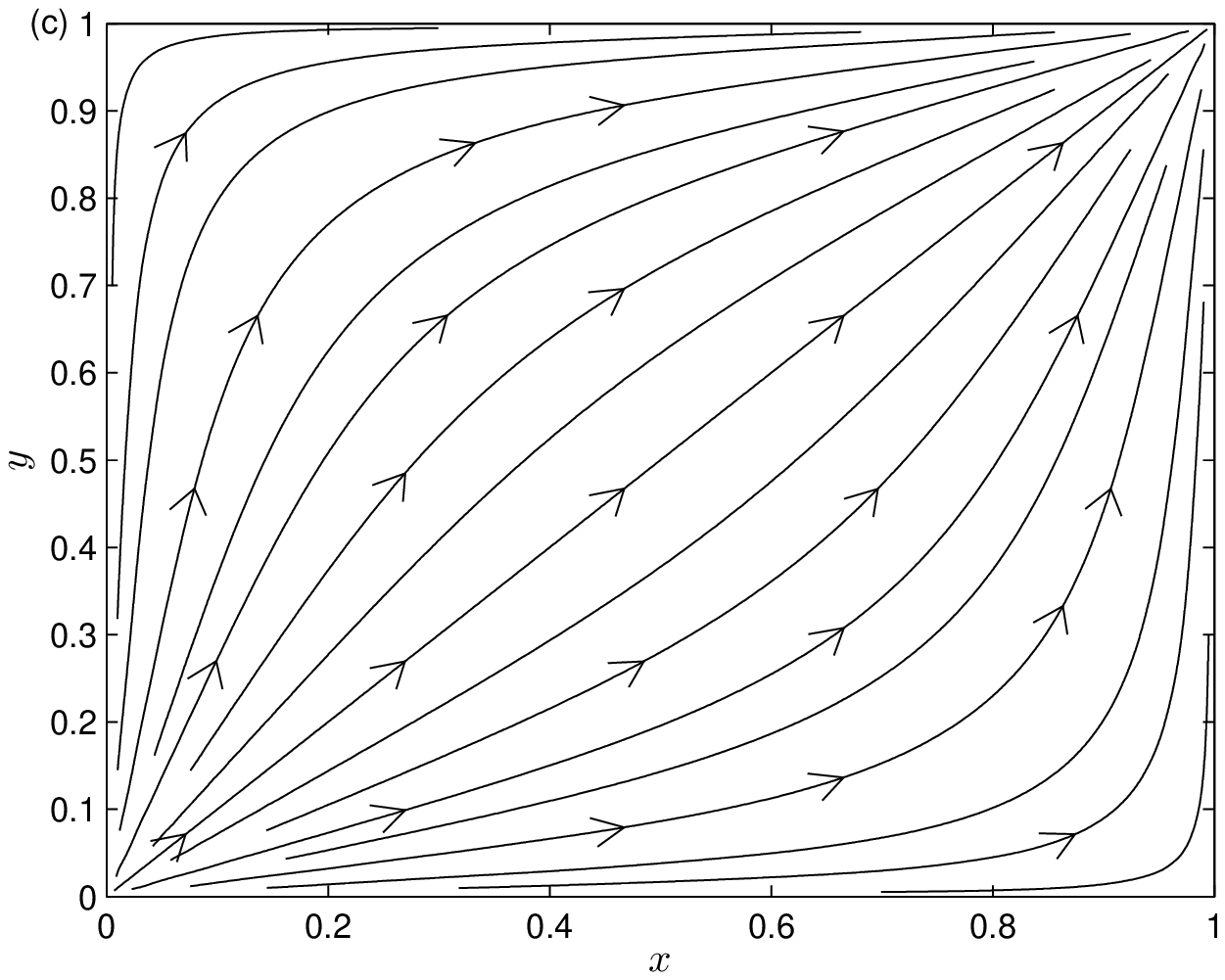}
\includegraphics[width=2.5in]{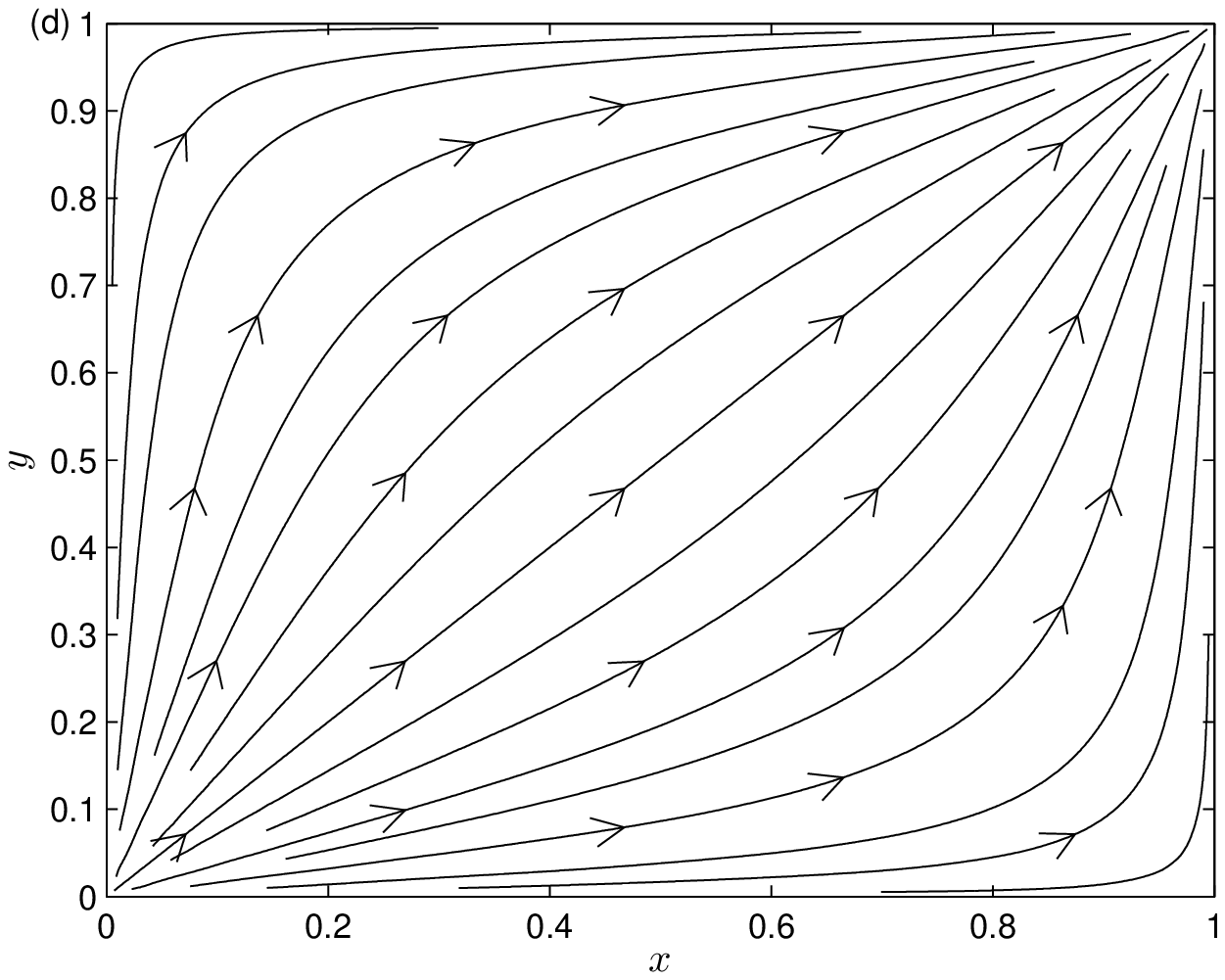}
\centering\caption{\label{fig:16} Streamlines at different time [(a): $T=10$, (b): $T=20$, (c): $T=30$, (d): $T=40$].}
\end{figure}

\section{Conclusions}

In this work, through properly constructing the equilibrium distribution functions, a LB model for two-phase flows in porous media is proposed at macroscopic scale level. With the Chapman-Enskog analysis, we can show that the macroscopic mathematical model for two-phase flows in porous media can be recovered correctly from present LB model. Then we also test present LB model with several classic problems, and find that the present results are in agreement with analytical solutions or some available numerical results. Finally, as an extension to the pore-scale LB models for multiphase flows in porous media, the present LB model is more suitable for large-scale problems governed by the macroscopic continuum models.

\section*{Acknowledgments}

This work was financially supported by the National Natural Science Foundation of China (Grants No. 51576079, No. 11602075 and No. 11602057).

\section*{Appendix A. Chapman-Enskog analysis of LB equation for the Poisson equation (\ref{eq2-15})}

In the Chapman-Enskog analysis, the distribution function $f_{i}(\mathbf{x},\;t')$, the derivatives of pseudo time and space, and the source term $F_{p}$ can be expanded as
\begin{subequations}\label{CEA1}
\begin{equation}
f_{i}=f_{i}^{(0)}+\epsilon f_{i}^{(1)}+\epsilon^{2} f_{i}^{(2)}+\cdots,
\end{equation}
\begin{equation}
 \frac{\partial}{\partial t'}=\epsilon\frac{\partial}{\partial t'_{1}}+\epsilon^{2}\frac{\partial}{\partial t'_{2}},\ \ \ \nabla=\epsilon \nabla_{1}, \ \ \ F_{p}=\epsilon^{2}F_{p}^{(2)}.
\end{equation}
\end{subequations}
Applying the Taylor expansion to Eq.~(\ref{eq3-1}), we have
\begin{equation}\label{TaylorA1}
\delta t D_{i} f_{i} + \frac{\delta t^{2}}{2} D_{i}^{2}f_{i} =  -\frac{1}{\tau_{f}}\big[f_{i} - f_{i}^{(eq)}\big]+\delta t\bar{\omega}_{i}F_{p},
\end{equation}
where $D_{i}=\frac{\partial}{\partial t'}+ \mathbf{c}_{i}\cdot\nabla$ is a differential operator.
Substituting Eq.~(\ref{CEA1}) into Eq.~(\ref{TaylorA1}), one can obtain the following equation,
\begin{eqnarray}
& &\big[\epsilon D_{i1}+\epsilon^{2}\frac{\partial}{\partial t'_{2}}+\frac{\delta t}{2}\big(\epsilon D_{i1}+\epsilon^{2}\frac{\partial}{\partial t'_{2}})^{2}\big](f_{i}^{(0)}+\epsilon f_{i}^{(1)}+\epsilon^{2} f_{i}^{(2)}+\cdots\big) \\ \nonumber
& = & -\frac{1}{\tau_{f}\delta t}\big(f_{i}^{(0)}- f_{i}^{(eq)}+\epsilon f_{i}^{(1)}+\epsilon^{2} f_{i}^{(2)}+\cdots\big)+\epsilon^{2}\bar{\omega}_{i}F_{p}^{(2)},
\end{eqnarray}
where $D_{i1}=\frac{\partial}{\partial t'_{1}}+ \mathbf{c}_{i}\cdot\nabla_{1}$. From above equation, we can also derive the zeroth, first and second-order equations in $\epsilon$,
\begin{subequations}
\begin{equation}\label{ConA1}
\epsilon^{0}: \ \ \ \ f_{i}^{(0)}=f_{i}^{(eq)},
\end{equation}
\begin{equation}\label{ConA2}
\epsilon^{1}: \ \ \ \ D_{i1}f_{i}^{(0)}=-\frac{1}{\tau_{f}\delta t}f_{i}^{(1)},
\end{equation}
\begin{equation}\label{ConA3}
\epsilon^{2}: \ \ \ \ \frac{\partial f_{i}^{(0)}}{\partial t'_{2}}+D_{i1}f_{i}^{(1)}+\frac{\delta t}{2}D_{i1}^{2}f_{i}^{(0)}=-\frac{1}{\tau_{f}\delta t}f_{i}^{(2)}+\bar{\omega}_{i}F_{p}^{(2)}.
\end{equation}
\end{subequations}
If we multiply the operator $D_{i1}$ on the both sides of Eq.~(\ref{ConA2}), and substitute the result into Eq.~(\ref{ConA3}), one can obtain
\begin{equation}\label{ConA33}
\epsilon^{2}: \ \ \ \ \frac{\partial f_{i}^{(0)}}{\partial t'_{2}}+D_{i1}\big(1-\frac{1}{2\tau_{f}}\big)f_{i}^{(1)}=-\frac{1}{\tau_{f}\delta t}f_{i}^{(2)}+\bar{\omega}_{i}F_{p}^{(2)}.
\end{equation}
In addition, based on Eqs.~ (\ref{eq3-7}a), (\ref{ConA1}) and (\ref{ConA2}), we can also derive the following equations,
\begin{equation}
\sum_{i}f_{i}^{(k)}=0,\ \ \ k \geq 1,
\end{equation}
\begin{eqnarray}\label{Cif1A1}
\sum_{i}\mathbf{c}_{i}f_{i}^{(1)} & = & -\tau_{f}\delta t \sum_{i}D_{i1}\mathbf{c}_{i}f_{i}^{(0)}=-\tau_{f}\delta t \sum_{i}D_{i1}\mathbf{c}_{i}f_{i}^{(eq)}\\ \nonumber
& = & -\tau_{f}\delta t c_{s}^{2}\nabla_{1} P,
\end{eqnarray}

After a summation of Eq.~(\ref{ConA33}), one can obtain the recovered equation at second-order of $\epsilon$,
\begin{equation}\label{ReA1}
\nabla_{1}\cdot\big[c_{s}^{2}\big(\frac{1}{2}-\tau_{f}\big)\delta t\nabla_{1}P\big]=F_{p}^{(2)}.
\end{equation}
Multiply $\epsilon^{2}$ on both sides of Eq.~(\ref{ReA1}), we can derive PE (\ref{eq2-15}) with $D_{p}$ determined by Eq.~(\ref{eq3-11}a).

We would also like to point out that following the idea in Refs. \cite{Chai2013,Chai2014,Meng2016}, the pressure gradient can be computed locally in the framework of LB method. Actually, from Eq.~(\ref{Cif1A1}) we have
\begin{equation}\label{Cif1A11}
\sum_{i}\mathbf{c}_{i}\epsilon f_{i}^{(1)} =  -\tau_{f}\delta t c_{s}^{2}\nabla P.
\end{equation}
Under assumption $\epsilon f_{i}^{(1)}\thickapprox f_{i}-f_{i}^{(eq)}$, we can derive the \emph{local} scheme [Eq.~(\ref{eq3-10}a)] for pressure gradient, which can also be used to compute the velocity [Eq.~(\ref{eq3-10}b)].

\section*{Appendix B. Chapman-Enskog analysis of LB equation for the convection-diffusion equation (\ref{eq2-18})}

Similar to above procedure, the distribution function $g_{i}(\mathbf{x},\;t)$, the derivatives of time and space, and the source term $F_{s}$ are expanded as
\begin{subequations}\label{CEB1}
\begin{equation}
g_{i}=g_{i}^{(0)}+\epsilon g_{i}^{(1)}+\epsilon^{2} g_{i}^{(2)}+\cdots,
\end{equation}
\begin{equation}
 \frac{\partial}{\partial t}=\epsilon\frac{\partial}{\partial t_{1}}+\epsilon^{2}\frac{\partial}{\partial t_{2}},\ \ \ \nabla=\epsilon \nabla_{1}, \ \ \ F_{s}=\epsilon F_{s}^{(1)}.
\end{equation}
\end{subequations}
Taking the Taylor expansion to Eq.~(\ref{eq3-2}), one can obtain
\begin{eqnarray}\label{TaylorB1}
\delta t \bar{D}_{i} g_{i} + \frac{\delta t^{2}}{2} \bar{D}_{i}^{2}g_{i} & = & -\frac{1}{\tau_{g}}\big[g_{i} - g_{i}^{(eq)}\big]+\delta t\omega_{i}\big(1-\frac{1}{2\tau_{g}}\big)F_{s}\nonumber \\
& + & \gamma\delta t\omega_{i}\frac{\mathbf{c}_{i}\cdot\nabla P}{\tau_{g}},
\end{eqnarray}
where $\bar{D}_{i}=\frac{\partial}{\partial t}+ \mathbf{c}_{i}\cdot\nabla$.

Substituting Eq.~(\ref{CEB1}) into Eq.~(\ref{TaylorB1}) yields
\begin{eqnarray}\label{CEB2}
& &\big[\epsilon \bar{D}_{i1}+\epsilon^{2}\frac{\partial}{\partial t_{2}}+\frac{\delta t}{2}\big(\epsilon \bar{D}_{i1}+\epsilon^{2}\frac{\partial}{\partial t_{2}}\big)^{2}\big]\big(g_{i}^{(0)}+\epsilon g_{i}^{(1)}+\epsilon^{2} g_{i}^{(2)}+\cdots\big) \\ \nonumber
& = & -\frac{1}{\tau_{g}\delta t}\big(g_{i}^{(0)}- g_{i}^{(eq)}+\epsilon g_{i}^{(1)}+\epsilon^{2} g_{i}^{(2)}+\cdots\big)+\epsilon\omega_{i}\big(1-\frac{1}{2\tau_{g}}\big) F_{s}^{(1)} + \epsilon \gamma\omega_{i}\frac{\mathbf{c}_{i}\cdot\nabla_{1} P}{\tau_{g}},
\end{eqnarray}
where $\bar{D}_{i1}=\frac{\partial}{\partial t_{1}}+ \mathbf{c}_{i}\cdot\nabla_{1}$. Based on Eq.~(\ref{CEB2}), we can also derive the zeroth, first and second-order equations in $\epsilon$,
\begin{subequations}
\begin{equation}\label{ConB1}
\epsilon^{0}: \ \ \ \ g_{i}^{(0)}=g_{i}^{(eq)},
\end{equation}
\begin{equation}\label{ConB2}
\epsilon^{1}: \ \ \ \ \bar{D}_{i1}g_{i}^{(0)}=-\frac{1}{\tau_{g}\delta t}g_{i}^{(1)}+\omega_{i}\big(1-\frac{1}{2\tau_{g}}\big) F_{s}^{(1)}+\gamma\omega_{i}\frac{\mathbf{c}_{i}\cdot\nabla_{1} P}{\tau_{g}},
\end{equation}
\begin{equation}\label{ConB3}
\epsilon^{2}: \ \ \ \ \frac{\partial g_{i}^{(0)}}{\partial t_{2}}+\bar{D}_{i1}g_{i}^{(1)}+\frac{\delta t}{2}\bar{D}_{i1}^{2}g_{i}^{(0)}=-\frac{1}{\tau_{g}\delta t}g_{i}^{(2)}.
\end{equation}
\end{subequations}
Multiplying the differential operator $\bar{D}_{i1}$ on the both sides of Eq.~(\ref{ConB2}), and substituting the result into Eq.~(\ref{ConB3}), we can rewrite the second-order equation in $\epsilon$ as
\begin{equation}\label{ConB33}
\epsilon^{2}: \ \ \ \ \frac{\partial g_{i}^{(0)}}{\partial t_{2}}+\bar{D}_{i1}\big(1-\frac{1}{2\tau_{g}}\big)\big[g_{i}^{(1)}+\frac{\delta t}{2} \omega_{i}F_{s}^{(1)}\big]+\frac{\delta t}{2}\bar{D}_{i1}\big(\gamma\omega_{i}\frac{\mathbf{c}_{i}\cdot\nabla_{1} P}{\tau_{g}}\big)=-\frac{1}{\tau_{g}\delta t}g_{i}^{(2)}.
\end{equation}
Besides, from Eqs.~(\ref{eq3-7}b), (\ref{eq3-9}b), (\ref{ConB1}) and (\ref{ConB2}), we can also obtain the following moments of non-equilibrium distribution function $g_{i}^{(1)}$,
\begin{equation}\label{Sgi}
\sum_{i}g_{i}^{(1)}=-\frac{\delta t}{2} F_{s}^{(1)},
\end{equation}
\begin{equation}\label{Sgik}
\sum_{i}g_{i}^{(k)}=0, \ \ \ k>1,
\end{equation}
\begin{eqnarray}\label{CigiB1}
\sum_{i}\mathbf{c}_{i}g_{i}^{(1)} & = & -\tau_{g}\delta t \sum_{i}\mathbf{c}_{i}\big[\bar{D}_{i1}g_{i}^{(0)}-\omega_{i}\big(1-\frac{1}{2\tau_{g}}\big) F_{s}^{(1)}-\gamma\omega_{i}\frac{\mathbf{c}_{i}\cdot\nabla_{1} P}{\tau_{g}}\big]\\ \nonumber
& = & -\beta\tau_{g}\delta t c_{s}^{2}\nabla_{1} S_{w}+\delta t \gamma c_{s}^{2}\nabla_{1} P.
\end{eqnarray}
With the help of Eqs.~(\ref{eq3-7}b) and (\ref{Sgi}), the recovered equation at first-order of $\epsilon$ can be derived through summing Eq.~(\ref{ConB2}) over $i$,
\begin{equation}\label{ReB1}
\phi\frac{\partial S_{w}}{\partial t_{1}}=F_{s}^{(1)}.
\end{equation}
After a summation of Eq.~(\ref{ConB33}), we can also obtain the recovered equation at second-order of $\epsilon$,
\begin{equation}\label{ReB2}
\phi\frac{\partial S_{w}}{\partial t_{2}}+\nabla_{1}\cdot\big[\beta c_{s}^{2}\big(\frac{1}{2}-\tau_{g}\big)\delta t\nabla_{1}S_{w}+c_{s}^{2}\delta t \gamma\nabla_{1}P\big]=0,
\end{equation}
where Eqs.~(\ref{eq3-7}b), (\ref{Sgi}), (\ref{Sgik}) and (\ref{CigiB1}) have been applied.

Multiplying  $\epsilon$ and $\epsilon^{2}$ on both sides of Eqs.~(\ref{ReB1}) and (\ref{ReB2}), we can derive equation (\ref{eq2-18}), and determine the parameter $A$ and diffusion coefficient $D_{s}$ through Eqs.~(\ref{eq3-8}) and (\ref{eq3-11}b).


\begin{thebibliography}{}
\bibitem{Bear1972} J. Bear, Dynamics of Fluids in Porous Media, Dover, New York, 1972.

\bibitem{Chen2006} Z. Chen, G. Huan, Y. Ma, Computaional Methods for Multiphase Flows in Porous Media, SIAM, Philadelphia, 2006.

\bibitem{Wu2016} Y.-S. Wu, Multiphase Fluid Flow in Porous and Fractured Reservoirs, Elsevier, Oxford, 2016.

\bibitem{Zhao2009} T. S. Zhao, C. Xu, R. Chen, W. W. Yang, Mass transport phenomena in direct methanol fuel cells, Prog. Eng. Combust. Sci. 35 (2009) 275-292.

\bibitem{Anderson1998} D. M. Anderson, G. B. McFadden, Diffuse-interface methods in fluid mechanics, Annu. Rev. Fluid Mech. 30 (1998) 139-165.

\bibitem{Tryggvason2011} G. Tryggvason, R. Scardovelli, S. Zaleski, Direct numerical simulations of gas-liquid multiphase flows, Cambridge University Press, Cambridge, 2011.

\bibitem{Blunt2013} M. J. Blunt, B. Bijeljic, H. Dong, O. Gharbi, S. Iglauer, P. Mostaghimi, A. Paluszny, C. Pentland, Pore-scale imaging and modelling, Adv. Water Resour. 51 (2013) 197-216.

\bibitem{Liu2016} H. Liu, Q. Kang, C. R. Leonardi, S. Schmieschek, A. Narv\'{a}ez, B. D. Jones, J. R. Williams, A. J. Valocchi, J. Harting, Multiphase lattice Boltzmann simulations for porous media applications, Comput. Geosci. 20 (2016) 777-805.

\bibitem{Chavent1986} G. Chavent, J. Jaffr\'{e}, Mathematical Models and Finite Elements for Reservoir Simulation, Elsevier, Amsterdam, 1986.

\bibitem{Jr1983} J. Douglas, Jr., Finite-difference methods for two-phase incompressible flow in porous media, SIAM J. Numer. Anal. 20 (1983) 681-696.

\bibitem{Enchery2006} G. Ench\'{e}ry, R. Eymard, A. Michel, Numerical approximation of a two-phase flow problem in a porous medium with discontinuous capillary forces, SIAM J. Numer. Anal. 43 (2006) 2402-2422.

\bibitem{Durlofsky2007} L. J. Durlofsky, Y. Efendiev, V. Ginting, An adaptive local¨Cglobal multiscale finite volume element method for two-phase flow simulations, Adv. Water Resour. 30 (2007) 576-588.

\bibitem{Lewis1984} R. N. Lewis, K. Morgan, K. H. Johnson, A finite element study of two-dimensional multiphase flow with particular reference to the five-spot problem, Comput. Meth. Appl. Mech. Eng. 44 (1984) 17-47.

\bibitem{Kukreti1989} A. R. Kukreti, Y. Rajapaksa, A numerical model for simulating two-phase flow through porous media, Appl. Math. Model. 13 (1989) 268-281.

\bibitem{Cao2011} Y. Cao, R. Helming, B. Wohlmuth, A two-scale operator-splitting method for two-phase flow in porous media, Adv. Water Resour. 34 (2011) 1581-1596.

\bibitem{Chueh2013} C.-H. Chueh, N. Djilali, W. Bangerth, An $h$-adaptive operator splitting method for two-phase flow in 3D heterogeneous porous media, SIAM J. Sci. Comput. 35 (2013) B149-B175.

\bibitem{Chen2004} Z. Chen, G. Huan, B. Li, An improved IMPES method for two-phase flow in porous media, Transp. Porous Med. 54 (2004) 361-376.

\bibitem{Yang2016} H. Yang, C. Yang, S. Sun, Active-set reduced-space methods with nonlinear elimination for two-phase flow problems in porous media, SIAM J. Sci. Comput. 38 (2016) B593-B618

\bibitem{Gerritsen2005} M. G. Gerritsen, L. J. Durlofsky, Modeling fluid flow in oil reservoirs, Annu. Rev. Fluid Mech. 37 (2005) 211-238

\bibitem{Chen1998}
 S. Chen, G. Doolen, Lattice Boltzmann method for fluid flows, Annu. Rev. Fluid Mech. 30 (1998) 329-364.

\bibitem{Succi2001} S. Succi, The Lattice Boltzmann Equation for Fluid Dynamics and Beyond, Oxford
University Press, Oxford, 2001.

\bibitem{Guo2013}
 Z. Guo, C. Shu, Lattice Boltzmann Method and Its Applications in Engineering, World Scientific, Singapore, 2013.

\bibitem{Kruger2017} T. Kr\"{u}ger, H. Kusumaatmaja, A. Kuzmin, O. Shardt, G. Silva, E. M. Viggen, The Lattice Boltzmann Method: Principles and Practice, Springer, Switzerland, 2017.

\bibitem{Chai2016} Z. Chai, B. Shi, Z. Guo, A multiple-relaxation-time lattice Boltzmann model for general nonlinear anisotropic convection-diffusion equations, J. Sci. Comput. 69 (2016) 355-390.

\bibitem{Chen2014} L. Chen, Q. Kang, Y Mu, Y.-L. He, W.-Q. Tao, A critical review of the pseudopotential multiphase lattice Boltzmann model: Methods and applications, Int. J. Heat Mass Transfer 76 (2014) 210¨C236.

\bibitem{Li2016} Q. Li, K.H. Luo, Q.J. Kang, Y.L. He, Q. Chen, Q. Liu, Lattice Boltzmann methods for multiphase flow and phase-change heat transfer, Prog. Energy Combust. Sci. 52 (2016) 62-105.

\bibitem{Xu2017} A. Xu, W. Shyy, T. S. Zhao, Lattice Boltzmann modeling of transport phenomena in fuel cells and flow batteries, Acta Mech. Sin. 33 (2017) 555-574.

\bibitem{Bultreys2016} T. Bultreys, W. De Boever, V. Cnudde, Imaging and image-based fluid transport modeling at the pore scale in geological materials: A practical introduction to the current state-of-the-art, Earth-Sci. Rev. 155 (2016) 93¨C128.

\bibitem{Qian1992}
Y. H. Qian, D. d'Humi\`{e}res, P. Lallemand, Lattice BGK models for Navier-Stokes equation, Europhys. Lett. 17 (1992) 479-484.

\bibitem{Ansumali2002} S. Ansumali, I. V. Karlin, Single relaxation time model for entropic lattice Boltzmann methods, Phys. Rev. E  65 (2002) 056312.

\bibitem{Ansumali2003} S. Ansumali, I. V. Karlin, H. C. \"{O}ttinger, Minimal entropic kinetic models for hydrodynamics, Europhys. Lett. 63 (2003) 798-804.

\bibitem{Ginzburg2005a}
I. Ginzburg, Equilibrium-type and link-type lattice Boltzmann models for generic advection and anisotropic-dispersion equation, Adv. Water Resour. 28 (2005) 1171-1195.

\bibitem{Ginzburg2008} I. Ginzburg, F. Verhaeghe, D. d'Humi\`{e}res, Two-relaxation-time lattice Boltzmann scheme: About parametrization, velocity, pressure and mixed boundary conditions, Commun. Comput. Phys. 3 (2008) 427-478.

\bibitem{dHumieres1992}
D. d'Humi\`{e}res, Generalized lattice-Boltzmann equations, in: B.D. Shizgal, D.P. Weave (Eds.), Rarefied Gas Dynamics: Theory and Simulations, in: Prog.
Astronaut. Aeronaut., Vol. 159, AIAA, Washington, DC, 1992, pp. 450-458.

\bibitem{Lallemand2000}
 P. Lallemand, L.-S. Luo, Theory of the lattice Boltzmann method: Dispersion, dissipation, isotropy, Galilean invariance, and stability, Phys. Rev. E 61 (2000) 6546-6562.

\bibitem{Geier2006} M. Geier, A. Greiner, J. G. Korvink, Cascaded digital lattice Boltzmann automata for high Reynolds number flow, Phys. Rev. E 73 (2006) 066705.

\bibitem{Premnath2011} K. N. Premnath, S. Banerjee, On the three-dimensional central moment lattice Boltzmann method, J. Stat. Phys. 143 (2011) 747-794.

\bibitem{He2000} X. He, N. Ling, Lattice Boltzmann simulation of electrochemical systems, Comput. Phys. Commun. 129 (2000) 158-166.

\bibitem{Hirabayashi2001} M. Hirabayashi, Y. Chen, H. Ohashi, The lattice BGK model for the Poisson equation, JSME Int. J. Ser. B 44 (2001) 45-52.

\bibitem{Wang2006} J. Wang, M. Wang, Z. Li, Lattice Poisson-Boltzmann simulations of electro-osmotic flows in microchannels, J. Colloid. Interface Sci. 296 (2006) 729-736.

\bibitem{Wang2011} H. Wang, G. Yan, B. Yan, Lattice Boltzmann model based on the Rebuilding-Divergency method for the Laplace equation and the Poisson equation, J. Sci. Comput. 46 (2011) 470-484.

\bibitem{Chai2008} Z. Chai, B. Shi, A novel lattice Boltzmann model for the Poisson equation, Appl. Math. Model., 32 (2008) 2050-2058.

\bibitem{Meng2016} X. Meng, Z. Guo, Localized lattice Boltzmann equation model for simulating miscible viscous displacement in porous media, Int. J. Heat Mass Transfer 100 (2016) 767-778.

\bibitem{Wolf-Gladrow1995} D. Wolf-Gladrow, A lattice Boltzmann model for diffusion, J. Stat. Phys. 79 (1995) 1023-1032.

\bibitem{Dawson1993} S. P. Dawson, S. Chen, G. Doolen, Lattice Boltzmann computations for reaction-diffusion equations,
J. Chem. Phys. 98 (1993) 1514-1523.

\bibitem{Shi2009} B. Shi, Z. Guo, Lattice Boltzmann model for nonlinear convection-diffusion equations, Phys. Rev. E 79 (2009) 016701.

\bibitem{Chopard2009} B. Chopard, J. L. Falcone, J. Latt, The lattice Boltzmann advection-diffusion model revisited, Eur. Phys. J. Special Topics 171 (2009) 245-249.

\bibitem{Huber2010} C. Huber, B. Chopard, M. Manga, A lattice Boltzmann model for coupled diffusion, J. Comput. Phys. 229 (2010) 7956-7976.

\bibitem{Yoshida2010}
H. Yoshida, M. Nagaoka, Multiple-relaxation-time lattice Boltzmann model for the convection and anisotropic diffusion equation, J. Comput. Phys. 229 (2010) 7774-7795.

\bibitem{Du2013} R. Du, W. Liu, A new multiple-relaxation-time lattice Boltzmann method for natural convection, J. Sci. Comput. 56 (2013) 122-130.

\bibitem{Huang2014} R. Huang, H. Wu, A modified multiple-relaxation-time lattice Boltzmann model for convection-diffusion equation, J. Comput. Phys. 274 (2014) 50-63.

\bibitem{Chai2013} Z. Chai, T. S. Zhao, Lattice Boltzmann model for the convection-diffusion equation, Phys. Rev. E 87 (2013) 063309.

\bibitem{Chai2014}
Z. Chai, T. S. Zhao, Nonequilibrium scheme for computing the flux of the convection-diffusion equation in the framework of the lattice Boltzmann method, Phys. Rev. E 90 (2014) 013305.

\bibitem{Yang2014} X. Yang, B. Shi, Z. Chai, Z. Guo, A coupled lattice Boltzmann method to solve Nernst-Planck model for simulating electro-osmotic flows, J. Sci. Comput. 61 (2014) 222-238.

\bibitem{Aursjo2017} O. Aursj{\o}, E. Jettestuen, J. L. Vinningland, A. Hiorth, An improved lattice Boltzmann method for simulating advective-diffusive processes in fluids, J. Comput. Phys. 332 (2017)363-375.

\bibitem{Li2017} L. Li, R. Mei, J. F. Klausner, Lattice Boltzmann models for the convection-diffusion equation: D2Q5 vs D2Q9, Int. J. Heat Mass Transfer 108 (2017) 41-62.

\bibitem{Ginzburg2005b}
I. Ginzburg, Generic boundary conditions for lattice Boltzmann models and their application to advection and anisotropic dispersion equations, Adv. Water Resour. 28 (2005) 1196-1216.

\bibitem{Zhang2012} T. Zhang, B. Shi, Z. Guo, Z. Chai, J. Lu, General bounce-back scheme for concentration boundary condition in the lattice-Boltzmann method, Phys. Rev. E 85 (2012) 016701.

\bibitem{Li2013} L. Li, R. Mei, J.F. Klausner, Boundary conditions for thermal lattice Boltzmann equation method, J. Comput. Phys. 237 (2013) 366-395.

\bibitem{Yong2015} J. Huang, W.-A. Yong, Boundary conditions of the lattice Boltzmann method for convection-diffusion equations, J. Comput. Phys. 300 (2015) 70-91.

\bibitem{Yong2017} W. Zhao, W.-A. Yong, Single-node second-order boundary schemes for the lattice Boltzmann method, J. Comput. Phys. 329 (2017) 1-15.

\bibitem{Cui2016} S. Cui, N. Hong, B. Shi, Z. Chai, Discrete effect on the halfway bounce-back boundary condition of multiple-relaxation-time lattice Boltzmann model for convection-diffusion equations, Phys. Rev. E 93 (2016) 043311.

\bibitem{Chai2016b} Z. Chai, C. Huang, B. Shi, Z. Guo, A comparative study on the lattice Boltzmann models for predicting effective diffusivity of porous media, Int. J. Heat Mass Transfer 98 (2016) 687-696.

\bibitem{Chen1999} Z. Chen, N. L. Khlopina, Degenerate two-phase incompressible flow problems III: Perturbation analysis and numerical experiments, Elect. J. Diff. Equ. 2 (1999) 29-49.

\bibitem{Ohlberger1997} M. Ohlberger, Convergence of a mixed finite elements-finite volume method for the two phase flow in porous media, East-West J. Numer. Math. 5(3) (1997) 183-210.

\bibitem{Mozolevski2013} I. Mozolevski, L. Schuh, Numerical simulations of two-phase immiscible incompressible flows in heterogeneous porous media with capillary barriers, J. Comput. Appl. Math. 242 (2013) 12-27.


\end{thebibliography}
\end{document}